\documentclass{LMCS}
\usepackage{graphicx}

\usepackage{url} 
\usepackage{enumerate,stmaryrd,mathrsfs}
\usepackage{amsmath,amssymb,latexsym}
\usepackage{hyperref}

\newcounter{zaehler_enumerate_kurz}

\newcommand*{\twodots}{.\,.\,}

\newcommand*{\einhalb}{\ensuremath{{\textstyle\frac{1}{2}}}}

\newcommand*{\raus}[1]{}

\newcommand*{\parno}{\par\noindent}

\newcommand*{\bs}[1]{\ensuremath{\boldsymbol{#1}}}

\newcommand*{\und}{\ensuremath{\wedge}}

\newcommand*{\oder}{\ensuremath{\vee}}

\newcommand*{\nicht}{\ensuremath{\neg}}
\newcommand*{\impl}{\ensuremath{\rightarrow}}

\renewcommand*{\geq}{\ensuremath{\geqslant}}
\renewcommand*{\leq}{\ensuremath{\leqslant}}

\newcommand*{\set}[1]{\ensuremath{\{ #1 \}}}
\newcommand*{\setc}[2]{\set{#1 \,:\, #2}}

\newcommand*{\bigset}[1]{\ensuremath{ \left\{ #1 \right\} } }

\newcommand*{\Pot}{\ensuremath{{\mathcal{P}}}}
\newcommand*{\dist}{\ensuremath{\textit{diff}}}
\newcommand*{\otype}{\ensuremath{{<}\textit{-type}}}

\newcommand*{\abgerundet}[1]{\ensuremath{
    \left\lfloor {#1} \right\rfloor }}

\newcommand*{\deff}{:=}

\newcommand*{\NN}{\ensuremath{\mathbb{N}}}    

\newcommand*{\A}{\ensuremath{\mathcal{A}}}
\newcommand*{\B}{\ensuremath{\mathcal{B}}}
\newcommand*{\C}{\ensuremath{\mathcal{C}}}

\newcommand*{\I}{\ensuremath{\mathcal{I}}}
\newcommand*{\J}{\ensuremath{\mathcal{J}}}

\newcommand*{\U}{\ensuremath{\mathcal{U}}}
\newcommand*{\bigO}{\ensuremath{{\mathcal{O}}}}

\newfont{\MyScript}{cmfi10 at 10pt}

\newcommand*{\struc}[1]{\ensuremath{\langle #1 \rangle}}

\newcommand*{\praedikat}[1]{\ensuremath{{\textsl{#1}}}}

\newcommand*{\Max}{\praedikat{max}}
\newcommand*{\Min}{\praedikat{min}}
\newcommand*{\Succ}{\praedikat{succ}}

\newcommand*{\Log}{\ensuremath{\lg}}

\newcommand*{\MIN}{\praedikat{MIN}\;}
\newcommand*{\MAX}{\praedikat{MAX}\;}

\newcommand*{\class}[1]{\ensuremath{\textrm{\upshape{#1}}}}

\newcommand*{\FO}{\class{FO}}

\newcommand*{\MSO}{\class{MSO}}

\newcommand*{\Fertig}{{}\hfill \ensuremath{\square}}
\newcommand*{\Qed} {\null\hfill\ensuremath{\blacksquare}}
 
\newcommand*{\FormelFont}[1]{\class{#1}}

\renewcommand*{\MSO}{\FormelFont{MSO}}
\newcommand*{\MLFP}{\FormelFont{MLFP}}

\newcommand*{\size}[1]{\ensuremath{||{#1}||}}

\newcommand*{\Tower}{\ensuremath{\textit{Tower}}}

\newcommand*{\BIT}{\ensuremath{\textup{bit}}}
\newcommand*{\BIN}{\ensuremath{\textup{BIN}}}

\newcommand*{\Tag}[1]{\ensuremath{\texttt{<#1>}}}
\newcommand*{\TagE}[1]{\ensuremath{\texttt{</#1>}}}

\newcommand*{\MonSigma}{\ensuremath{\FormelFont{Mon}\Sigma}}

\newcommand*{\Il}{\ensuremath{\textit{il}}}
\newcommand*{\Fl}{\ensuremath{\textit{sl}}}

\newcommand*{\T}{{\ensuremath{\mathcal{T}}}}

\newcommand*{\ORD}[1]{\ensuremath{\textit{Ord}({#1})}}
\newcommand*{\DIST}[2]{\ensuremath{\textit{Dist}_{#1}(#2)}}
\newcommand*{\TYPE}[2]{\ensuremath{\textit{Type}_{#1}(#2)}}

\newcommand*{\tildelta}{\ensuremath{\tilde{\delta}}}

\usepackage{amsthm} 

\newcommand*{\fertig}{{}\hfill \ensuremath{\square}}
\renewcommand*{\qed} {\null\hfill\ensuremath{\blacksquare}}

\newtheorem{quasi_theorem}{Theorem}[section] 
\newtheorem*{quasi_theorem_nn}{Theorem} 

\newtheorem{quasi_lemma}[quasi_theorem]{Lemma}
\newtheorem*{quasi_lemma_nn}{Lemma} 

\newtheorem{quasi_corollary}[quasi_theorem]{Corollary}
\newtheorem*{quasi_corollary_nn}{Corollary} 

\newtheorem{quasi_proposition}[quasi_theorem]{Proposition}
\newtheorem*{quasi_proposition_nn}{Proposition} 

\newtheorem{quasi_maintheorem}[quasi_theorem]{Theorem}
\newtheorem*{quasi_maintheorem_nn}{Theorem} 

\newtheorem{quasi_definition}[quasi_theorem]{Definition}
\newtheorem*{quasi_definition_nn}{Definition} 

\newtheorem{quasi_example}[quasi_theorem]{Example}
\newtheorem*{quasi_example_nn}{Example} 

\newtheorem{quasi_remark}[quasi_theorem]{Remark}
\newtheorem*{quasi_remark_nn}{Remark} 

\newtheorem{quasi_remarks}[quasi_theorem]{Remarks}
\newtheorem*{quasi_remarks_nn}{Remarks} 

\newtheorem{quasi_question}[quasi_theorem]{Question}
\newtheorem*{quasi_question_nn}{Question} 

\newtheorem{quasi_facts}[quasi_theorem]{Facts}
\newtheorem*{quasi_facts_nn}{Facts} 

\newtheorem{quasi_claim}{Claim}
\newtheorem*{quasi_claim_nn}{Claim} 

\newtheorem{quasi_fact}{Fact}
\newtheorem*{quasi_fact_nn}{Fact} 

\newenvironment{proof_ohne}{{\it Proof:}}{}
\newenvironment{proofc_ohne}[1]{{\rm \bf Proof {#1}:}}{}
\newenvironment{proofsketch_ohne}{{\it Proof (sketch):}}{}

\newenvironment{theorem}{\begin{quasi_theorem}}{\end{quasi_theorem}\noindent}

\newenvironment{lemma}{\begin{quasi_lemma}}{\end{quasi_lemma}\noindent}

\newenvironment{corollary}{\begin{quasi_corollary}}{\end{quasi_corollary}\noindent}

\newenvironment{definition}{\begin{quasi_definition}\rm}{\end{quasi_definition}\noindent}

\newenvironment{example}{\begin{quasi_example}\rm}{\end{quasi_example}\noindent}

\newenvironment{remark}{\begin{quasi_remark}\rm}{\end{quasi_remark}\noindent}

\newenvironment{claim}{\begin{quasi_claim}}{\end{quasi_claim}\noindent}

\renewenvironment{proof}{\begin{proof_ohne}}{\end{proof_ohne}}
\newenvironment{proofc}[1]{\begin{proofc_ohne}{#1}}{\end{proofc_ohne}}

\newtheorem{quasi_Todo}[quasi_theorem]{To Do}

\usepackage[latin1]{inputenc}

\def\doi{1 (1:6) 2005}
\lmcsheading%
{\doi}
{25}
{}
{}
{Nov.~\phantom{0}4, 2004}
{Jun.~29, 2005}
{}
\begin{document}

\title[The succinctness of first-order logic on linear orders]{The succinctness of first-order logic on linear orders}

\author[M.\ Grohe]{Martin Grohe}     
\address{Institut f\"ur Informatik,
         Humboldt-Universit\"at,
         Unter den Linden 6, D-10099 Berlin, Germany}   
\email{\url{{grohe | schweika}@informatik.hu-berlin.de}}  

\author[N.\ Schweikardt]{Nicole Schweikardt} 



\begin{abstract}
Succinctness is a natural measure for comparing the strength of different
logics. Intuitively, a logic $L_1$ is more succinct than another logic $L_2$
if all properties that can be expressed in $L_2$ can be expressed in $L_1$ by
formulas of (approximately) the same size, but some properties can be
expressed in $L_1$ by (significantly) smaller formulas.

We study the succinctness of logics on linear orders.
Our first theorem is concerned with the finite variable fragments of
first-order logic. We prove that:
\begin{enumerate}
\item[(i)] Up to a polynomial factor, the 2- and the 3-variable fragments
  of first-order logic on linear orders have the same succinctness.
\item[(ii)]
The 4-variable fragment is exponentially more succinct than the
  3-variable fragment.
\end{enumerate}

Our second main result compares the succinctness of first-order logic on
linear orders with that of monadic second-order logic.
We prove that the fragment of monadic second-order logic that has the same expressiveness as
first-order logic on linear orders is non-elementarily more succinct than
first-order logic.
\end{abstract}

\maketitle
\section{Introduction}\label{section:Introduction}%

It is one of the fundamental themes of logic in computer science to study and
compare the \emph{strength} of various logics. Maybe the most natural measure
of strength is the \emph{expressive power} of a logic. By now, researchers
from finite model theory, but also from more application driven areas such as
database theory and automated verification, have developed a rich toolkit that has
led to a good understanding of the expressive power of the fundamental logics
(e.g.\ \cite{EbbinghausFlum,Immerman,Libkin_FMT-textbook}). It should also be said that there are clear
limits to the understanding of expressive power, which are often linked to
open problems in complexity theory.

In several interesting situations, however, one encounters different logics
of the same expressive power. As an example, let us consider node selecting
query languages for XML-documents. Here the natural deductive query language
monadic datalog \cite{gotkoc04} and various automata based query ``languages''
\cite{nev99,nevschwe99,frigrokoc03} have the same expressive power as monadic
second-order logic. XML-documents are usually modelled by labelled trees.
Logics on trees and strings also play an important role in automated
verification. Of the logics studied in the context of
verification, the modal $\mu$-calculus is another logic that has the same
expressive power as monadic second-order logic on ranked trees and strings,
and linear time temporal logic LTL has the same expressive power as
first-order logic on strings \cite{kam68}. 

\emph{Succinctness} is a natural
measure for comparing the strength of logics that have the same expressive
power. Intuitively, a logic $L_1$ is more succinct than another logic $L_2$
if all properties that can be expressed in $L_2$ can be expressed in $L_1$ by
formulas of (approximately) the same size, but some properties can be
expressed in $L_1$ by (significantly) smaller formulas.

For both expressiveness and succinctness there is a trade-off between the
strength of a logic and the complexity of evaluating formulas of the logic.
The difference lies in the way the complexity is measured. Expressiveness is
related to \emph{data complexity}, which only takes into account the size of
the structure in which the formula has to be evaluated, whereas succinctness is
related to the \emph{combined complexity}, which takes into account both the
size of the formula and the structure \cite{var82}.

Succinctness has received surprisingly little attention so far. A few
scattered results are \cite{wil99,aleimm00,adlimm01,etevarwil02,sto74}; for example, it
is known that first-order
logic on strings is non-elementarily more succinct than LTL \cite{kam68,sto74}. In
\cite{GroheSchweikardt_CSL03}, we started a more systematic investigation. Specifically,
we studied the succinctness of various logics on trees that all have the same
expressive power as monadic second-order logic. While we were able to gain a
reasonable picture of the succinctness of these logics, it also became clear
that we are far from a thorough understanding of succinctness. In particular,
very few techniques for proving lower bounds are available.

Most of the lower bound proofs use automata theoretic arguments, often
combined with a clever encoding of large natural numbers that goes back to
Stockmeyer \cite{sto74}. In \cite{GroheSchweikardt_CSL03}, these techniques were
also combined with complexity theoretic reductions to prove lower bounds on
succinctness under certain complexity theoretic assumptions. Wilke
\cite{wil99} used refined automata theoretic arguments to prove that
$\text{CTL}^+$ is exponentially more succinct than $\text{CTL}$. Adler and
Immerman~\cite{adlimm01} were able to improve Wilke's lower bound slightly,
but what is more important is that they introduced games for establishing
lower bounds on succinctness. These games vaguely resemble
Ehrenfeucht-Fra\"{\i}ss\'e games, which are probably the most important tools
for establishing inexpressibility results. 

In this paper, we study the succinctness of logics on linear orders
(without any additional structure).
In particular, we consider finite variable fragments of
first-order logic. It is known and easy to see that even the 2-variable
fragment has the same expressive power as full first-order logic on linear
orders (with respect to Boolean and unary queries). 
We prove the following theorem:
\begin{theorem}\label{theo:mt1}
\begin{enumerate}
\item[(i)] Up to a polynomial factor, the 2 and the 3-variable fragments of
  first-order logic on linear orders have the same succinctness.
\item[(ii)]
The 4-variable fragment of
  first-order logic on linear orders is exponentially more succinct than the
  3-variable fragment.\Fertig
\end{enumerate}
\end{theorem}%
For the sake of completeness, let us also mention that full first-order logic
is at most exponentially more succinct than the 3-variable fragment.  It
remains an open problem if there is also an exponential gap in succinctness
between full first-order logic and the 4-variable fragment.

Of course the main result here is the exponential gap in
Theorem~\ref{theo:mt1} (ii), but it should be noted that (i) is also by no
means obvious.  The theorem may seem very technical and not very impressive at
first sight, but we believe that to gain a deeper understanding of the issue
of succinctness it is of fundamental importance to master basic problems such
as those we consider here first (similar, maybe, to basic inexpressibility
results such as the inability of first-order logic to express that a linear
order has even length). The main technical result behind both parts of the
theorem is that a 3-variable first-order formula stating that a linear order
has length $m$ must have size at least $\frac{1}{2}\sqrt{m}$. Our technique
for proving this result originated in the Adler-Immerman games, even though
later it turned out that the proofs are clearer if the reference to the game
is dropped. 

There is another reason the gap in succinctness between the 3- and 4-variable
fragments is interesting: It is a long standing open problem in finite model
theory if, for $k\ge 3$, the $k$ variable fragment of first-order logic is
strictly less expressive than the $(k{+}1)$-variable fragment on the class
of all ordered finite structures. 
This question is still open for all $k\ge 3$. Our result (ii) at
least shows that there are properties that require exponentially larger
3-variable than 4-variable formulas.

Succinctness as a measure for comparing the strength of logics is not
restricted to logics of the same expressive power. Even if a logic $L_1$ is
more expressive than a logic $L_2$, it is interesting to know whether
those properties that can be expressed in both $L_1$ and $L_2$ can be
expressed more succinctly in one of the logics. Sometimes, this may even be
more important than the fact that some esoteric property is expressible in
$L_1$, but not $L_2$. We compare first-order logic with the more expressive
monadic second-order logic and prove:
\begin{theorem}\label{theo:mt2}
The fragment of monadic second-order logic that has the same expressiveness as
first-order logic on linear orders is non-elementarily more succinct than
first-order logic.~\Fertig
\end{theorem}%

The paper is organised as follows: After the Preliminaries, in
Section~\ref{section:LowerBoundProof} we prove the main technical result
behind Theorem~\ref{theo:mt1}. In Sections~\ref{section:FO2vsFO3} and
\ref{section:FOvsFO3} we formally state and prove the two parts of the theorem.
Finally, Section~\ref{section:FOvsMSO} is devoted to Theorem~\ref{theo:mt2}.

The present paper is the full version of the conference contribution \cite{GS_LICS04}.

\section{Preliminaries}\label{section:Preliminaries}%
We write $\NN$ for the set of non-negative integers.
\\
We assume that the reader is familiar with first-order logic $\FO$
(cf., e.g., the textbooks \cite{EbbinghausFlum,Immerman}).
For a natural number $k$ we write $\FO^k$ to denote the \emph{$k$-variable fragment}
of $\FO$. 
The three variables available in $\FO^3$ will always be denoted $x$, $y$, and $z$.
We write $\FO^3(<,\Succ,\Min,\Max)$ (resp., $\FO^3(<)$) to denote the class of all $\FO^3$-formulas of
signature $\set{<,\Succ,\Min,\Max}$ (resp., of signature $\set{<}$), with binary relation symbols 
$<$ and $\Succ$ and constant symbols $\Min$ and $\Max$.
In the present paper, such formulas will always be interpreted in finite structures where $<$ is a linear
ordering, $\Succ$ the successor relation associated with $<$, and $\Min$ and $\Max$ the minimum and maximum
elements w.r.t.\ $<$.
\par
For every $N\in\NN$ let $\A_N$ be the $\set{<,\Succ,\Min,\Max}$-structure
with universe $\set{0,\twodots,N}$, $<$ the natural linear ordering, 
$\Min^{\A_N} = 0$, $\Max^{\A_N} = N$, and
$\Succ$ the relation with $(a,b)\in\Succ$ iff $a{+}1=b$. 
We identify the \emph{class of linear orders} with the set $\setc{\A_N}{N\in \NN}$.
\par
For a structure $\A$ we write $\U^{\A}$ to denote $\A$'s universe.
When considering $\FO^3$, an \emph{interpretation} is a tuple $(\A,\alpha)$, where
$\A$ is one of the structures $\A_N$ (for some $N\in\NN$) and 
$\alpha : \set{x,y,z}\to \U^{\A}$ is a variable assignment in $\A$.
To simplify notation, we will extend every assignment $\alpha$ to a 
mapping $\alpha : \set{x,y,z,\Min,\Max}\rightarrow \U^{\A}$,
letting $\alpha(\Min) = \Min^{\A}$ and
$\alpha(\Max) = \Max^{\A}$.
For a variable $v\in \set{x,y,z}$ and an element $a\in\U^{\A}$ we
write $\alpha[\frac{a}{v}]$ to denote the assignment that maps $v$ to $a$ and that coincides with $\alpha$ on
all other variables.
If $A$ is a set of interpretations and $\varphi$ is an $\FO^3(<,\Succ,\Min,\Max)$-formula, we write
$A\models\varphi$ to indicate that $\varphi$ is satisfied by \emph{every} interpretation in $A$.
\par
In a natural way, we view formulas as finite trees (precisely, as their \emph{syntax trees}), where leaves
correspond to the atoms of the formulas, and inner vertices correspond
to Boolean connectives or quantifiers.  We
define the \emph{size} $\size{\varphi}$ of $\varphi$ to be
the number of vertices of $\varphi$'s syntax tree.
\begin{definition}[Succinctness]\mbox{}\\
Let $L_1$ and $L_2$ be logics, let $F$ be a class of functions from $\NN$ to $\NN$, and let
$\C$ be a class of structures.
We say that \emph{$L_1$ is $F$-succinct in $L_2$ on $\C$} 
iff there is a function $f\in F$ such
that for every
$L_1$-sentence $\varphi_1$ there is an $L_2$-sentence $\varphi_2$ of size 
$\size{\varphi_2}\leq f(\size{\varphi_1})$ which is equivalent to $\varphi_1$ on all
structures in $\C$.\Fertig 
\end{definition}%
Intuitively, a logic $L_1$ being
$F$-succinct in a logic $L_2$  means that $F$ gives
an upper bound on the size of $L_2$-formulas needed to express
\emph{all} of $L_1$. This definition may seem slightly at odds with
the common use of the
term ``succinctness'' in
statements such as ``$L_1$ is exponentially \emph{more succinct} than
$L_2$'' meaning that there is \emph{some} $L_1$-formula that is not
equivalent to any $L_2$-formula of sub-exponential size. In our
terminology we would rephrase this last statement as ``$L_1$ is \emph{not}
$2^{o(m)}$-succinct in $L_2$'' (here we interpret sub-exponential as
$2^{o(m)}$, but of course this is not the issue). The reason for
defining $F$-succinctness the way we did is that it makes the formal
statements of our results much more convenient. We will continue to use
statements such as ``$L_1$ is exponentially more succinct than $L_2$''
in informal discussions.
\begin{example}
$\FO^3(<,\Succ,\Min,\Max)$ is $\bigO(m)$-succinct in $\FO^3(<)$ on the class of linear orders,
because $\Succ(x,y)$ (respectively, $x{=}\Min$, respectively, $x{=}\Max$) can be expressed by the formula
\[
   (x<y) \ \  \und \ \ \nicht \, \exists\, z\; \big( (x<z)\  \und \ (z < y)\big)
\] 
(respectively, $\nicht \exists y\, (y<x)$, \ respectively, 
$\nicht \exists y \,(x<y)$).
\Fertig
\end{example}%
\section{Lower bound for $\FO^3$}\label{section:LowerBoundProof}%
\subsection{Lower Bound Theorem}%
%
Before stating our main lower bound theorem, we need some more 
notation.
\par
If $S$ is a set we write $\Pot_2(S)$ for the set of all 
2-element subsets of $S$.
For a finite subset $S$ of $\NN$ we write $\MAX S$ (respectively, $\MIN S$) to denote
the maximum (respectively, minimum) element in $S$.
For integers $m,n$ we define 
\begin{eqnarray*}
 \dist(m,n) & \deff &  m-n
\end{eqnarray*}
to be the difference between $m$ and $n$.
We define \,$\otype(m,n)\in\set{<,=,>}$\, as follows: 
\begin{center}
\begin{tabular}{lcll}
if & $m<n$ & then & $\otype(m,n)$ \ $\deff$ \ $\mbox{``${<}$''}$\,,\\
if & $m=n$ & then & $\otype(m,n)$ \ $\deff$ \ $\mbox{``${=}$''}$\,,\\
if & $m>n$ & then & $\otype(m,n)$ \ $\deff$ \ $\mbox{``${>}$''}$.
\end{tabular}
\end{center}
We next fix the notion of a \emph{separator}. Basically, if $A$ and
$B$ are sets of interpretations and $\delta$ is a separator for $\struc{A,B}$,
then $\delta$ contains information that allows to distinguish every 
interpretation $\I\in A$ from every interpretation $\J\in B$.
\begin{definition}[separator]\label{definition:separator} \mbox{}\\
Let $A$ and $B$ be sets of interpretations.\\
A \emph{potential separator} is a mapping
\[
  \delta \ : \ \Pot_2\big(\set{\Min,\Max,x,y,z}\big)\longrightarrow \NN\,.
\]
$\delta$ is called \emph{separator for $\struc{A,B}$}, if the following is
satisfied:
For every $\I\deff(\A,\alpha)\in A$ and $\J\deff(\B,\beta)\in B$ there are
$u,u' \in \set{\Min,\Max,x,y,z}$ with $u\neq u'$, such that
\,$\delta\big(\set{u,u'}\big)\geq 1$ \ and
\begin{enumerate}[1.]
\item
  $\otype\big(\alpha(u),\alpha(u')\big) \ \ \neq \ \ \otype\big(\beta(u),\beta(u')\big)$
  \quad or
\item
  {
   $\delta\big(\set{u,u'}\big) \geq 
   \MIN\big\{ |\dist\big(\alpha(u),\alpha(u')\big)|\,,\ 
              |\dist\big(\beta(u),\beta(u')  \big)| 
       \big\}$
  }%
  \quad and \\
  $\dist\big(\alpha(u),\alpha(u')\big) \ \neq \ \dist\big(\beta(u),\beta(u')\big)$. 
  \fertig
\end{enumerate}
\end{definition}%
Note that $\delta$ is a separator for $\struc{A,B}$ if, and only if, $\delta$ is
a separator for $\struc{\set{\I},\set{\J}}$, for all $\I\in A$ and $\J\in B$.
For simplicity, we will often write $\struc{\I,\J}$ instead of $\struc{\set{\I},\set{\J}}$.
\medskip\\
Let us now state an easy lemma on the existence of separators.
\begin{lemma}\label{lemma:existence_separators}
If $A$ and $B$ are sets of interpretations for which there exists an
$\FO^3(<)$-formula $\psi$ such that $A\models\psi$ and
$B\models\!\nicht\psi$, then there exists a \emph{separator}
$\delta$ for $\struc{A,B}$.\fertig
\end{lemma}%
\begin{proof} 
We need the following notation:
For a number $d\in\NN$ and an interpretation $(\A,\alpha)$ choose
\[
  \ORD{\A,\alpha} \  : \ \set{0,1,2,3,4}\rightarrow \set{\Min,x,y,z,\Max}
\]
such that
$\alpha\big(\ORD{\A,\alpha}(i)\big) \leq
 \alpha\big(\ORD{\A,\alpha}({i+1})\big)
$, for all $0\leq i < 4$. 
Furthermore, choose
\[
  \DIST{d}{\A,\alpha} \ : \ \setc{\,(i,{i+1})}{0\leq i < 4}
                 \rightarrow 
                 \set{0,\twodots,2^{d+1}}
\]
such that the following is true for all $0\leq i < 4$:
{
\begin{eqnarray*}
  \DIST{d}{\A,\alpha}(i,{i+1}) & = &
   \dist\Big( \alpha\big( \ORD{\A,\alpha}(i{+}1) \big),
             \alpha\big( \ORD{\A,\alpha}({i}) \big)
       \Big),\quad\mbox{or}
  \\
  \DIST{d}{\A,\alpha}(i,{i+1}) \ = \ 2^{d+1} & \leq &
  \dist\Big( \alpha\big( \ORD{\A,\alpha}(i{+}1) \big),
             \alpha\big( \ORD{\A,\alpha}({i}) \big)
       \Big).
\end{eqnarray*}
}%
Finally, we define the \emph{$d$-type} of $(\A,\alpha)$ as
\begin{eqnarray*}
  \TYPE{d}{\A,\alpha} & \deff & \big(\,\ORD{\A,\alpha},\ \DIST{d}{\A,\alpha}\,\big).
\end{eqnarray*}
Using an Ehrenfeucht-Fra\"\i{}ss\'{e} game, it is an easy exercise to
show the following (cf., e.g., \cite{EbbinghausFlum}):
\begin{lemma}\label{lemma:d-Type}
Let $d\in\NN$ and let $(\A,\alpha)$ and $(\B,\beta)$ be interpretations. 
If $\TYPE{d}{\A,\alpha} = \TYPE{d}{\B,\beta}$, then 
$(\A,\alpha)$ and $(\B,\beta)$ cannot be distinguished by $\FO(<)$-formulas of 
quantifier depth $\leq d$.~\fertig
\end{lemma}%
Let $d$ be the quantifier depth of the formula $\psi$. We define $\delta$ to be
the potential separator
with $\delta(p)\deff 2^{d+1}$, for all 
$p\in\Pot_2(\set{\Min,\Max,x,y,z})$.
\\
To show that $\delta$ is, in fact, a \emph{separator} for $\struc{A,B}$, let
$(\A,\alpha)\in A$ and $(\B,\beta)\in B$. Since $(\A,\alpha)\models\psi$ and
$(\B,\beta)\not\models\psi$, we obtain from Lemma~\ref{lemma:d-Type} that
$\TYPE{d}{(\A,\alpha)}\neq \TYPE{d}{(\B,\beta)}$, i.e., 
\begin{enumerate}[1.]
 \item
  $\ORD{\A,\alpha}\neq \ORD{\B,\beta}$, \quad or
 \item
  $\DIST{d}{\A,\alpha} \neq \DIST{d}{\B,\beta}$.
\end{enumerate}
Therefore, there exist 
$u,u' \in \set{\Min,\Max,x,y,z}$ with $u\neq u'$, such that
\begin{enumerate}[1.]
\item
  $\otype\big(\alpha(u),\alpha(u')\big) \ \ \neq \ \ \otype\big(\beta(u),\beta(u')\big)$,
  \quad or
\item
  {
   $\dist\big(\alpha(u),\alpha(u')\big) \ \neq \ \dist\big(\beta(u),\beta(u')\big)$
   \quad and  \\
   $\delta\big(\set{u,u'}\big) = 2^{d+1} \geq
    \MIN\big\{ |\dist\big(\alpha(u),\alpha(u')\big)|\,,\ 
               |\dist\big(\beta(u),\beta(u')  \big)| 
       \big\}$.
  }%
\end{enumerate}
Consequently, $\delta$ is a separator for $\struc{A,B}$, and
the proof of Lemma~\ref{lemma:existence_separators} is complete.~\qed
\end{proof}
%
\begin{definition}\label{definition:weight}[weight of $\delta$]\mbox{}\\
Let $\delta$ be a potential separator. We define 
\begin{enumerate}[(a)]
\item
 the \emph{border-distance} 
  \begin{eqnarray*}
    b(\delta) & \deff &
    \MAX\; \big\{\ \delta(\set{\Min, \Max}), \ \   
              \delta(\set{\Min,u})+\delta(\set{u',\Max}) \ : \ 
              u,u'\in\set{x,y,z}\ \big\}
  \end{eqnarray*}
\item
 the \emph{centre-distance} 
 \begin{eqnarray*}
   c(\delta) & \deff &
   \MAX \;
    \big\{\ \delta(p)+\delta(q) \ : \ 
     p,q\in\Pot_2(\set{x,y,z}),\ p\neq q\ \big\}
 \end{eqnarray*}
\item
 the \emph{weight}
  \begin{eqnarray*}
    w(\delta) & \deff & \sqrt{c(\delta)^2 + b(\delta)}.
  \end{eqnarray*}
\vspace{-6ex}\\ \mbox{}\fertig
\end{enumerate}
\end{definition}%
There is not much intuition we can give for this particular choice of 
weight function, except for the fact that it seems to be exactly
what is needed for the proof of our main lower bound theorem (Theorem~\ref{theorem:lower_bound}).
At least it will later, in Remark~\ref{remark:choice_of_weight}, become clear why
the $\sqrt{\ \,}$-function is used for defining the weight function.
\begin{definition}[minimal separator] \mbox{}\\
$\delta$ is called a \emph{minimal separator} for $\struc{A,B}$ if
$\delta$ is a separator for $\struc{A,B}$ and
\begin{eqnarray*}
 w(\delta) & = & \MIN\big\{ \,w(\delta') \,:\, 
 \delta' \mbox{ is a separator for } \struc{A,B} \,\big\}.
\end{eqnarray*}
\vspace{-6ex}\\ \mbox{}\fertig
\end{definition}%
Now we are ready to formally state our main lower bound theorem on the
size of $\FO^3(<)$-formulas:
\begin{theorem}\label{theorem:lower_bound}{\upshape [main lower bound theorem]}\mbox{}\\
If $\psi$ is an $\FO^3(<,\Succ,\Min,\Max)$-formula, $A$ and $B$ are
sets of interpretations such that $A\models \psi$ and
$B\models\nicht\psi$, and
$\delta$ is a minimal separator for $\struc{A,B}$, then
\begin{eqnarray*}
  \size{\psi} & \geq & \einhalb\cdot {w(\delta)}\,. 
\end{eqnarray*}
\vspace{-6ex}\\ \mbox{}\fertig
\end{theorem}%
Before giving details on the proof of Theorem~\ref{theorem:lower_bound}, let us first
point out its following easy consequence:
\begin{corollary}\label{cor:lower_bound}\mbox{}\\
Let $n>m\geq 0$. The two linear orders $\A_m$ and $\A_n$ (with universe $\set{0,\twodots,m}$ and
$\set{0,\twodots,n}$, respectively) cannot be distinguished by
an $\FO^3(<,\allowbreak\Succ,\Min,\Max)$-sentence of size 
$< \einhalb\sqrt{m}$. \mbox{}\fertig
\end{corollary}%
\begin{proof}
Let $\psi$ be an $\FO^3(<,\Succ,\Min,\Max)$-sentence with
$\A_m\models\psi$ and $\A_n\models\nicht\psi$. Let $\alpha$ be the assignment that
maps each of the variables $x$, $y$, and $z$ to the value $0$.
Consider the
mapping 
\[
 \delta_m \ : \ \Pot_2(\set{\Min,\Max,x,y,z})\rightarrow \NN
 \qquad\mbox{defined via}
\]
\[
 \delta_m(p) \ \deff \ \left\{
   \begin{array}{lll}
     m & , & \mbox{if \ } p=\set{\Min,\Max} \\
     0 & , & \mbox{otherwise.}
   \end{array}
 \right.
\]
It is straightforward to check that
\ $w(\delta_m) = \sqrt{m}$ \ and that
$\delta_m$ is a \emph{minimal separator} for 
\[
  \struc{\,(\A_m,\alpha)\,,\,(\A_n,\alpha)\,}\,.
\] 
From Theorem~\ref{theorem:lower_bound} we therefore obtain that
\[
 \size{\psi} \quad \geq \quad \einhalb\cdot{w(\delta_m)} \quad = \quad
 \einhalb\cdot \sqrt{m}.
\]
This completes the proof of Corollary~\ref{cor:lower_bound}.\mbox{}\qed
\end{proof}%
\\
\parno
To prove Theorem~\ref{theorem:lower_bound} we need a series of 
intermediate results, as well
as the notion of an \emph{extended syntax tree} of a formula, which
is a syntax tree where each node carries an additional label containing
information about sets of interpretations satisfying, respectively, not satisfying, 
the associated subformula.
More precisely, every node $v$ of the extended syntax tree carries 
an \emph{interpretation label} $\Il(v)$
which consists of a pair $\struc{A,B}$ of sets of interpretations such that every interpretation in
$A$, but no interpretation in $B$, satisfies the subformula represented by the subtree rooted
at node $v$.
Basically, such an extended syntax tree corresponds to a {game tree} that is constructed
by the two players of the \emph{Adler-Immerman game} (cf., \cite{adlimm01}).
\par
For proving Theorem~\ref{theorem:lower_bound}
we consider an extended syntax tree $\T$ of the given formula $\psi$.
We define a weight function on the nodes of $\T$ by defining the weight
$w(v)$ of each node $v$ of $\T$ to be the weight of a \emph{minimal
separator} for $\Il(v)$. Afterwards --- and this is the main technical 
difficulty --- we show that the weight of each node $v$ is bounded (from above) by 
the weights of $v$'s children. This, in turn, enables us to prove a 
lower bound on the number of nodes in $\T$ which depends on
the weight of the root node.
\subsection{Proof of Theorem~\ref{theorem:lower_bound}}
We start with the formal definition of extended syntax trees.
\begin{definition}\label{definition:extended_syntax_tree}
[extended syntax tree]\mbox{}\\
Let $\psi$ be an $\FO^3(<,\Succ,\Min,\Max)$-formula,
let $A$ and $B$ be sets of interpretations such that $A\models\psi$ and
$B\models\nicht\psi$. By induction on the construction of $\psi$ we define an
\emph{extended syntax tree} $\T_{\psi}^{\struc{A,B}}$ as follows:
\begin{enumerate}[$\bullet$]
\item
 If $\psi$ is an atomic formula, then $\T_{\psi}^{\struc{A,B}}$
 consists of a single node $v$ that has a \emph{syntax label} $\Fl(v) \deff \psi$ and
 an \emph{interpretation label} $\Il(v) \deff \struc{A,B}$.
\item
 If $\psi$ is of the form $\nicht\psi_1$, then 
 $\T_{\psi}^{\struc{A,B}}$ has 
 a root node $v$ with $\Fl(v) \deff \nicht$\, and \,$\Il(v)\deff \struc{A,B}$.
 The unique child of $v$ is the root of $\T_{\psi_1}^{\struc{B,A}}$.
 Note that $B\models\psi_1$ and $A\models\nicht\psi_1$.
\item
 If $\psi$ is of the form $\psi_1\oder\psi_2$, then $\T_{\psi}^{\struc{A,B}}$ has 
 a root node $v$ with $\Fl(v) \deff \oder$\, and \,$\Il(v)\deff \struc{A,B}$.
 \\
 The first child of $v$
 is the root of $\T_{\psi_1}^{\struc{A_1,B}}$. The second child of $v$ is
 the root of $\T_{\psi_2}^{\struc{A_2,B}}$, where, for $i\in\set{1,2}$,
 $A_i = \setc{(\A,\alpha)\in A}{(\A,\alpha)\models \psi_i}$.
 \\
 Note that $A=A_1\cup A_2$, $A_i\models\psi_i$, and $B\models\nicht\psi_i$.
\item
 If $\psi$ is of the form $\psi_1\und\psi_2$, then $\T_{\psi}^{\struc{A,B}}$ has 
 a root node $v$ with $\Fl(v) \deff \und$\, and \,$\Il(v)\deff \struc{A,B}$.
 \\
 The first child of $v$
 is the root of $\T_{\psi_1}^{\struc{A,B_1}}$. The second child of $v$ is
 the root of $\T_{\psi_2}^{\struc{A,B_2}}$, where, for $i\in\set{1,2}$,
 $B_i = \setc{(\B,\beta)\in B}{(\B,\beta)\not\models \psi_i}$.
 \\
 Note that $B=B_1\cup B_2$, $A\models\psi_i$, and $B_i\models\nicht\psi_i$.
\item
 If $\psi$ is of the form $\exists u\,\psi_1$, for a variable
 $u\in\set{x,y,z}$, then $\T_{\psi}^{\struc{A,B}}$ has a
 root node $v$ with $\Fl(v) \deff \exists u$\, and \,$\Il(v)\deff \struc{A,B}$.
 The unique child of $v$
 is the root of $\T_{\psi_1}^{\struc{A_1,B_1}}$,
 where 
 $B_1 \deff \setc{(\B,\beta[\frac{b}{u}])}{(\B,\beta)\in B,\ b\in \U^{\B}}$, and
 $A_1$ is chosen as follows:
 For every $(\A,\alpha)\in A$ fix an element $a\in\U^{\A}$ such that
 $\big(\A,\alpha[\frac{a}{u}]\big) \models \psi_1$, and let
 $A_1 \deff \setc{(\A,\alpha[\frac{a}{u}])}{(\A,\alpha)\in A}$.
 \\
 Note that $A_1\models \psi_1$ and $B_1\models\nicht\psi_1$.
\item
 If $\psi$ is of the form $\forall u\,\psi_1$, for a variable
 $u\in\set{x,y,z}$, then $\T_{\psi}^{\struc{A,B}}$ has a
 root node $v$ with $\Fl(v) \deff \forall u$\, and \,$\Il(v)\deff \struc{A,B}$.
 The unique child of $v$
 is the root of $\T_{\psi_1}^{\struc{A_1,B_1}}$,
 where 
 $A_1 \deff \setc{(\A,\alpha[\frac{a}{u}])}{(\A,\alpha)\in A,\ a\in \U^{\A}}$, and
 $B_1$ is chosen as follows:
 For every $(\B,\beta)\in B$ fix an element $b\in\U^{\B}$ such that
 $\big(\B,\beta[\frac{b}{u}]\big) \models \nicht\psi_1$, and let
 $B_1 \deff \setc{(\B,\alpha[\frac{b}{u}])}{(\B,\beta)\in B}$.
 \\
 Note that $A_1\models \psi_1$ and $B_1\models\nicht\psi_1$.
 \fertig
\end{enumerate}
\end{definition}%
The following is the main technical result necessary for our proof of 
Theorem~\ref{theorem:lower_bound}.
\begin{lemma}\label{lemma:key-prop}
Let $\psi$ be an $\FO^3(<,\Succ,\Min,\Max)$-formula, let
$A$ and $B$ be sets of interpretations such that $A\models\psi$ and
$B\models\nicht\psi$, and
let $\T$ be an extended syntax tree $\T_{\psi}^{\struc{A,B}}$.
\\
For every node $v$ of $\T$ the following is true, where $\delta$ is a
minimal separator for $\Il(v)$:
\begin{enumerate}[(a)]
\item\label{key-prop:leaf}
 If $v$ is a \emph{leaf}, then \ $w(\delta) \leq 1$.
\item\label{key-prop:two}
 If $v$ has 2 children $v_1$ and $v_2$,
 and $\delta_i$ is a minimal separator for $\Il(v_i)$, for $i\in\set{1,2}$, then
 \\
 $
   w(\delta) \ \ \leq \ \ w(\delta_1) + w(\delta_2)\,.
 $%
\item\label{key-prop:one}
 If $v$ has exactly one child $v_1$, and 
 $\delta_1$ is a minimal separator for $\Il(v_1)$, then \ 
 $w(\delta) \leq  \ w(\delta_1)  +  2.$%
\fertig
\end{enumerate}%
\end{lemma}%
The proof of Lemma~\ref{lemma:key-prop} is given in 
Section~\ref{subsection:key_prop} below.%
\medskip\\
For a binary tree $\T$ we write $\size{\T}$ to denote the number of nodes of $\T$.
For the proof of Theorem~\ref{theorem:lower_bound} we also need the following 
easy observation.
\begin{lemma}\label{lemma:weighted_trees}
Let $\T$ be a finite binary tree where each node $v$ is equipped with a
\emph{weight} $w(v) \geq 0$ such that the following is true:
\begin{enumerate}[(a)]
\item
  If $v$ is a {leaf}, then $w(v) \leq 1$.
\item
  If $v$ has 2 children $v_1$ and $v_2$, then \ 
  $w(v) \ \leq \ w(v_1) + w(v_2)$.
\item
  If $v$ has exactly one child $v_1$, then \ 
  $w(v) \ \leq \  w(v_1) + 2$.  
\end{enumerate}
Then, 
\ $
  \size{\T} \ \geq \ \einhalb\cdot w(r)\,,
$ \ 
where $r$ is the root of $\T$.~\fertig
\end{lemma}%
\begin{proof} 
By induction on the size of $\T$.
\\
If $\T$ consists of a single node $v$, then $\size{\T} = 1 \geq \einhalb\cdot 1$;
and $1\geq w(v)$, since $v$ is a leaf.
\\
If $\T$ consists of a root node $v$ whose first child $v_1$ is the root of a tree $\T_1$ and
whose second child $v_2$ is the root of a tree $\T_2$, then $\size{\T} = 1 + \size{\T_1} + \size{T_2}$.
By induction we know for $i\in\set{1,2}$ that $\size{\T_i}\geq \einhalb {w(v_i)}$.
From the assumption we have that $w(v)\leq w(v_1)+w(v_2)$.
Therefore, 
\[
  \size{\T} \quad \geq \quad 
   1 + \einhalb {w(v_1)} + \einhalb {w(v_2)} \quad \geq \quad 
  \einhalb w(v).
\]
If $\T$ consists of a root node $v$ whose unique child $v_1$ is the root of a tree $\T_1$, then
$\size{\T} = 1 + \size{\T_1}$.
By induction we know that $\size{\T_1}\geq \einhalb {w(v_1)}$. 
From the assumption we have that $w(v)\leq  w(v_1) + 2$, i.e.,
$\einhalb w(v) \leq \einhalb w(v_1) + 1$. 
Therefore, $\size{\T} \geq 1 + \einhalb w(v_1) \geq \einhalb w(v)$.  
\\
This completes the proof of Lemma~\ref{lemma:weighted_trees}.
\qed
\end{proof}
\\
\parno
%
Using Lemma~\ref{lemma:key-prop} and~\ref{lemma:weighted_trees}, we
are ready for the
\\
\parno
\begin{proofc}{of Theorem~\ref{theorem:lower_bound}}\mbox{}\\
We are given an $\FO^3(<,\Succ,\Min,\Max)$-formula $\psi$ and sets $A$ and $B$ of interpretations
such that $A\models\psi$ and $B\models\nicht\psi$.
Let $\T$ be an extended syntax tree $\T_{\psi}^{\struc{A,B}}$.
\par
We equip each node $v$ of $\T$ with a \emph{weight} $w(v)\deff w(\delta_v)$, where
$\delta_v$ is a {minimal separator} for $\Il(v)$.
From Lemma~\ref{lemma:key-prop} we obtain that the preconditions of
Lemma~\ref{lemma:weighted_trees} are satisfied.
Therefore,
$\size{\T}\geq \einhalb \cdot w(r)$, where $r$ is the root of 
$\T$, i.e., $w(r) = w(\delta)$, for a minimal separator $\delta$
for $\Il(r) = \struc{A,B}$.
\\
From Definition~\ref{definition:extended_syntax_tree} it should be obvious that
$\size{\psi} = \size{\T}$. Therefore, the proof of Theorem~\ref{theorem:lower_bound}
is complete.
\qed
\end{proofc}%
%
%
%
\subsection{Proof of Lemma~\ref{lemma:key-prop}}\label{subsection:key_prop}
%
We partition the proof of Lemma~\ref{lemma:key-prop} into proofs for 
the parts \emph{(\ref{key-prop:leaf})}, \emph{(\ref{key-prop:two})}, and \emph{(\ref{key-prop:one})},
where part \emph{(\ref{key-prop:one})} turns out to be the most elaborate.
\par
According to the assumptions of Lemma~\ref{lemma:key-prop} we are given 
an $\FO^3(<,\Succ,\Min,\Max)$-formula $\psi$ and 
sets $A$ and $B$ of interpretations such that $A\models\psi$ and
$B\models\nicht\psi$.
Furthermore, we are given
an extended syntax tree $\T = \T_{\psi}^{\struc{A,B}}$.
Throughout the remainder of this section, $\T$ will always denote this particular syntax tree.
\\
\parno
%
%
\begin{proofc}{of part \emph{(\ref{key-prop:leaf})} of Lemma~\ref{lemma:key-prop}}\mbox{}\\
Let $v$ be a \emph{leaf} of $\T$ and let $\delta$ be a
minimal separator for $\struc{A_v,B_v} \deff \Il(v)$. 
Our aim is to show that $w(\delta) \leq 1$.
\par
By Definition~\ref{definition:extended_syntax_tree} we know that $\Fl(v)$ is an \emph{atomic} formula
of the form $R(u,u')$ for $R\in\set{{<},{=},\Succ}$ and $u,u'\in\set{\Min,\Max,x,y,z}$. 
Furthermore, $A_v\models R(u,u')$ and $B_v\models\nicht R(u,u')$.
I.e., for all $(\A,\alpha)\in A_v$ and $(\B,\beta)\in B_v$, 
\begin{eqnarray*}
 \otype\big(\alpha(u),\alpha(u')\big) & \neq & \otype\big(\beta(u),\beta(u')\big)
\\
  & \mbox{or} &
\\
 |\dist\big(\alpha(u),\alpha(u')\big)|\ \  = \ \ 1
 & \neq &
 |\dist\big(\beta(u),\beta(u')\big)|.
\end{eqnarray*} 
In case that $u\neq u'$ we can define a \emph{separator} $\tildelta$ for $\struc{A_v,B_v}$ 
via 
\begin{eqnarray*}
 \tildelta(p) & \deff  & 
 \left\{ \begin{array}{lll}
  1 & , & \mbox{if } \  p=\set{u,u'} \\
  0 & , & \mbox{otherwise.}
 \end{array}
 \right.
\end{eqnarray*}
Since $\delta$ is a \emph{minimal} separator, we obtain that
$w(\delta)\leq w(\tildelta) = 1$.
\par
It remains to consider the case where $u=u'$. Here, $A_v\models R(u,u)$ and $B_v\models \nicht R(u,u)$.
Since $R\in\set{{<},{=},\Succ}$ this implies that $A_v=\emptyset$ or $B_v=\emptyset$.
Therefore, according to Definition~\ref{definition:separator}, the mapping $\tildelta$ with
$\tildelta(p)=0$, for all $p\in\Pot_2(\set{\Min,\Max,x,y,z})$, is a {separator} for
$\struc{A_v,B_v}$. Hence, $w(\delta)\leq w(\tildelta) = 0$.
\\
This completes the proof of part \emph{(\ref{key-prop:leaf})} of Lemma~\ref{lemma:key-prop}.
\qed
\end{proofc}
%
\\
\parno
The essential step in the proof of 
part \emph{(\ref{key-prop:two})} of Lemma~\ref{lemma:key-prop} is
the following easy lemma.
\begin{lemma}\label{lemma:separator-two}
Let $v$ be a node of $\T$ that has two children $v_1$ and $v_2$. Let $\delta_1$ and
$\delta_2$ be separators for $\Il(v_1)$ and $\Il(v_2)$, respectively.
Let $\tildelta$ be the
{potential separator} defined on every $p\in\Pot_2(\set{\Min,\Max,x,y,z})$ via
\begin{eqnarray*}
   \tildelta(p) & \deff & \delta_1(p) + \delta_2(p)\,.
\end{eqnarray*}
Then, $\tildelta$ is a \emph{separator} for $\Il(v)$.~\fertig
\end{lemma}%
%
\begin{proof}
Let $\struc{A,B}\deff\Il(v)$.
We need to show that $\tildelta$ is a separator for 
$\struc{\I,\J}$, 
for all $\I\in A$ and $\J\in B$.
Let therefore
$\I\deff(\A,\alpha)\in A$ and $\J\deff(\B,\beta)\in B$ be 
fixed for the remainder of this proof. 
\par
Since $v$ has 2 children, we know from Definition~\ref{definition:extended_syntax_tree}
that \,$\Fl(v)= \oder$\, or \,$\Fl(v)= \und$.
Let us first consider the case where \,$\Fl(v)= \oder$.
\par
From Definition~\ref{definition:extended_syntax_tree} we know
that, for $i\in\set{1,2}$, $\Il(v_i) = \struc{A_i,B}$, where 
$A_1\cup A_2 = A$. 
Therefore, there is an $i\in\set{1,2}$ such that $\I\in A_i$.
From the assumption we know that $\delta_i$ is a separator for $\struc{A_i,B}$.
Therefore, there are 
$u,u' \in \set{\Min,\Max,x,y,z}$ with $u\neq u'$, such that
$\delta_i\big(\set{u,u'}\big)\geq 1$ and
\begin{enumerate}[1.]
\item
  $\otype\big(\alpha(u),\alpha(u')\big) \ \ \neq \ \ \otype\big(\beta(u),\beta(u')\big)$
  \quad or
\item
   $\delta\big(\set{u,u'}\big) \geq 
   \MIN\big\{ |\dist\big(\alpha(u),\alpha(u')\big)|\,,\ 
              |\dist\big(\beta(u),\beta(u')  \big)| 
       \big\}$
  \quad and \\
  $\dist\big(\alpha(u),\alpha(u')\big) \ \neq \ \dist\big(\beta(u),\beta(u')\big)$. 
\end{enumerate}
Since $\tildelta(\set{u,u'}) = \delta_1(\set{u,u'}) + \delta_2(\set{u,u'})$, we know that
$\tildelta(\set{u,u'}) \geq \delta_i(\set{u,u'})$. Therefore, $\tildelta$ is a 
separator for $\struc{\I,\J}$.
This completes the proof of Lemma~\ref{lemma:separator-two} for the case that
$\Fl(v) = \oder$.
\par
The case $\Fl(v) = \und$ follows by symmetry.
\qed
\end{proof}
%
%
\\
\parno
Using Lemma~\ref{lemma:separator-two}, the proof of 
part \emph{(\ref{key-prop:two})} of Lemma~\ref{lemma:key-prop} is straightforward:
\\
\parno
%
\begin{proofc}{of part \emph{(\ref{key-prop:two})} of Lemma~\ref{lemma:key-prop}}\mbox{}\\
Let $v$ be a node of $\T$ that has two children $v_1$ and $v_2$. Let $\delta$, $\delta_1$, and
$\delta_2$, respectively, be minimal separators for $\Il(v)$, $\Il(v_1)$, and $\Il(v_2)$, respectively.
Our aim is to show that $w(\delta) \leq w(\delta_1) + w(\delta_2)$.
\par
Let $\tildelta$ be the \emph{separator for $\Il(v)$} obtained from Lemma~\ref{lemma:separator-two}.
Since $\delta$ is a \emph{minimal} separator for $\Il(v)$, it suffices to show that
 \ $w(\tildelta) \ \leq \ w(\delta_1) + w(\delta_2)$.
\par
Using Definition~\ref{definition:weight}, it is straightforward
to check that $b(\tildelta)\leq b(\delta_1) + b(\delta_2)$ and
$c(\tildelta)\leq c(\delta_1) + c(\delta_2)$.
From this we obtain that
\begin{eqnarray*}
  w(\tildelta)^2
&  = 
& c(\tildelta)^2 + b(\tildelta) 
\\
&  \leq
& \big( c(\delta_1) + c(\delta_2) \big)^2 + b(\delta_1) + b(\delta_2)
\\
&  =
& c(\delta_1)^2 + b(\delta_1) + c(\delta_2)^2 + b(\delta_2) + 2c(\delta_1)c(\delta_2)
\\
&  \leq
& w(\delta_1)^2 + w(\delta_2)^2 + 2w(\delta_1)w(\delta_2)
\\
&  =
& \big( w(\delta_1) + w(\delta_2) \big)^2\,.
\end{eqnarray*}
I.e., we have shown that \ $w(\tildelta)\leq w(\delta_1) + w(\delta_2)$.
\\
This completes the proof of  
part \emph{(\ref{key-prop:two})} of Lemma~\ref{lemma:key-prop}.
\qed
\end{proofc}
%
\\
\parno
An essential step in the proof of 
part \emph{(\ref{key-prop:one})} of Lemma~\ref{lemma:key-prop} is
the following lemma.
\begin{lemma}\label{lemma:separator-one}
Let $v$ be a node of $\T$ that has 
syntax-label 
$\Fl(v) = \textsf{\upshape Q}u$, for
$\textsf{\upshape Q}\in \set{\exists,\forall}$ and
$ u\in\set{x,y,z}$.
Let $\delta_1$
be a separator for $\Il(v_1)$, where $v_1$ is the unique child of $v$ in $\T$.
Let $\tildelta$ be the {potential separator} defined via
{%
\begin{enumerate}[$\bullet$]
\item
$\tildelta(\set{u,u'}) \ \deff \ 0$\,,\quad 
for all \,$u'\in\set{\Min,\Max,x,y,z}\setminus\set{u}$\,,
\item
$\tildelta(\set{\Min,\Max})
  \ \deff \
  \MAX \big\{\;  
      \delta_1(\set{\Min,\Max})
      \,, \ \ 
      \delta_1(\set{\Min,u}) + \delta_1(\set{u,\Max}) + 1
    \;\big\}$\,, 
\end{enumerate}%
}
\noindent{}and for all $u',u''$ such that $\set{x,y,z}=\set{u,u',u''}$ and
all $m\in\set{\Min,\Max}$,
{%
\begin{enumerate}[$\bullet$]
\item
$\tildelta(\set{u',u''})
 \ \deff \
 \MAX \big\{\;
     \delta_1(\set{u',u''})\,,\ \  
     \delta_1(\set{u',u}) + \delta_1(\set{u,u''}) + 1
   \;\big\}$\,,
\item
$\tildelta(\set{m,u'})
 \ \deff \
 \MAX \big\{\;
     \delta_1(\set{m,u'})\,,\ \ 
     \delta_1(\set{m,u}) + \delta_1(\set{u,u'}) + 1
   \;\big\}$\,.
\end{enumerate}
}
\noindent{}Then, $\tildelta$ is a \emph{separator} for $\Il(v)$.~\fertig
\end{lemma}%
%
%
\begin{proof}
We only consider the case where $\textsf{\upshape Q} u = \exists z$. 
All other cases $\textsf{\upshape Q}\in \set{\exists,\forall}$ and $u\in \set{x,y,z}$ 
follow by symmetry.
\par
Let $\struc{A,B}\deff\Il(v)$.
We need to show that $\tildelta$ is a separator for 
$\struc{\I,\J}$, 
for all $\I\in A$ and $\J\in B$. Let therefore
$\I\deff(\A,\alpha)\in A$ and $\J\deff(\B,\beta)\in B$ be fixed for the remainder of this proof. The aim is to show that $\tildelta$ is a separator for $\struc{\I,\J}$.
\par
Since $\Fl(v)=\exists z$, we know from Definition~\ref{definition:extended_syntax_tree} that
$\Il(v_1) = \struc{A_1,B_1}$, where $B_1$ contains the interpretations 
$(\B,\beta[\frac{b}{z}])$, for all $b\in\U^{\B}$, and $A_1$ contains an interpretation
$(\A,\alpha[\frac{a}{z}])$, for a particular $a\in \U^{\A}$.
We define $\alpha_a\deff \alpha[\frac{a}{z}]$, $\I_a\deff (\A,\alpha_a)$, and for every
$b\in\U^{\B}$, $\beta_b\deff \beta[\frac{b}{z}]$ and $\J_b\deff (\B,\beta_b)$.
\par
From the fact that $\delta_1$ is a separator for $\Fl(v_1)$, we in particular know, for 
every $b\in\U^{\B}$, that
$\delta_1$ is a separator for $\struc{\I_a,\J_b}$. 
I.e., we know the following:
\medskip\\
For every $b\in\U^{\B}$ there are $u_b,u'_b \in\set{\Min,\Max,x,y,z}$ with \,$u_b\neq u'_b$,
such that
\emph{$\struc{\I_a,\J_b}$ is separated by $\delta_1(\set{u_b,u'_b})$}, i.e.,
\,$\delta_1(\set{u_b,u'_b})\geq 1$ \,and
{
\begin{enumerate}[(1)${}_a$:] 
\item[$(1)_b$:]
  {
  $\otype\big(\alpha_a(u_b),\alpha_a(u'_b)\big) \ \neq \ \otype\big(\beta_b(u_b),\beta_b(u'_b)\big)$,
  }%
  \quad or
\item[$(2)_b$:]
   {
   $\delta_1\big(\set{u_b,u'_b}\big) \ \geq \ 
    \MIN\big\{\, |\dist\big(\alpha_a(u_b),\alpha_a(u'_b)\big)|\,,\ \
               |\dist\big(\beta_b(u_b),\beta_b(u'_b)  \big)| 
       \,\big\}$
  }%
  \quad and \\
  {
  $\dist\big(\alpha_a(u_b),\alpha_a(u'_b)\big) \ \neq \ \dist\big(\beta_b(u_b),\beta_b(u'_b)\big)$. 
  }%
 \smallskip 
\end{enumerate}%
}
\noindent{}In what follows we will prove a series of claims which ensure 
that $\tildelta$ is a separator for $\struc{\I,\J}$. 
%
We start with 
\begin{claim}\label{claim:1}
If there is a \,$b\in\U^{\B}$ such that $\struc{\I_a,\J_b}$ is separated by
   $\delta_1(\set{u_b,u'_b})$ with $z\not\in\set{u_b,u'_b}$, 
  then $\tildelta$ is a separator for $\struc{\I,\J}$.~\fertig  
\end{claim}%
\begin{proof}
As $z\not\in\set{u_b,u'_b}$, we have, by definition of $\tildelta$,
that $\tildelta(\set{u_b,u'_b})\geq
\delta_1(\set{u_b,u'_b})$. Therefore, $(1)_b$ and $(2)_b$
imply that $\tildelta$ is a separator for $\struc{\I_a,\J_b}$ as well as for
$\struc{\I,\J}$.\\
This completes the proof of Claim~\ref{claim:1}.~\qed 
\end{proof}
\medskip\\ 
%
Due to Claim~\ref{claim:1} it henceforth suffices to assume that 
for no \,$b\in\U^{\B}$, $\struc{\I_a,\J_b}$ is separated by
   $\delta_1(\set{u_b,u'_b})$ with $z\not\in\set{u_b,u'_b}$.
I.e., we assume that for every $b\in\U^{\B}$, $\struc{\I_a,\J_b}$ is separated
by $\delta_1(\set{\Min,z})$, $\delta_1(\set{z,\Max})$, $\delta_1(\set{x,z})$, or
$\delta_1(\set{y,z})$.
%
\begin{claim}\label{claim:2}
If \,$a=\alpha(u)$\, for some $u\in\set{\Min,\Max,x,y}$, then
$\tildelta$ is a separator for $\struc{\I,\J}$.~\fertig
\end{claim}%
\begin{proof} 
Choose $b\deff \beta(u)$. Therefore, $\alpha_a(z) = a = \alpha(u) = \alpha_a(u)$ and
$\beta_b(z) = b = \beta(u) = \beta_b(u)$.
\par
We know that $\struc{\I_a,\J_b}$ is separated by 
$\delta_1(\set{z,u'})$, for some $u'\in\set{\Min,\Max,x,y}$.
Furthermore, since $\alpha_a(z) = \alpha_a(u)$ and $\beta_b(z) = \beta_b(u)$,
we have $u'\neq u$.
\par
By definition of $\tildelta$ we know that
$\tildelta(\set{u,u'}) \geq \delta_1(\set{z,u'})$.
Therefore, $\tildelta$ is a separator for 
$\struc{\I_a,\J_b}$ as well as for $\struc{\I,\J}$.
\\
This completes the proof of Claim~\ref{claim:2}.~\qed
\end{proof}
\medskip\\
%
Due to Claim~\ref{claim:2} it henceforth suffices to assume that, 
\,$a\neq \alpha(u)$, \,for
all $u\in\set{\Min,\Max,x,y}$.
\medskip

%
\begin{claim}~\label{claim:3}
If \,$\delta_1(\set{\Min,z})\geq \dist(a,\Min^{\A})$,\, then
$\tildelta$ is a separator for $\struc{\I,\J}$.~\fertig 
\end{claim}%
\begin{proof}
We distinguish between two cases.
An illustration is given in Figure~\ref{fig:Claim3}.
\begin{figure}
\begin{center}
  \includegraphics[width=12cm]{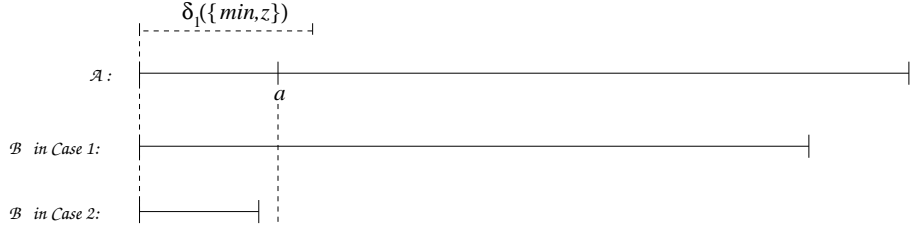}
\end{center}
\caption{Situation in Claim 3.}\label{fig:Claim3}
\end{figure}
\medskip\\
\textit{\underline{Case~1:} \ $\dist(\Max^{\B},\Min^{\B}) \geq \dist(a,\Min^{\A})$}.\\
In this case we can choose $b\in\U^{\B}$ with 
\begin{eqnarray*}
  \dist(b,\Min^{\B}) & = & \dist(a,\Min^{\A})
\end{eqnarray*} 
(simply via $b\deff a$).
Obviously, $\struc{\I_a,\J_b}$ is not separated by $\delta_1(\set{\Min,z})$.
However, we know that $\struc{\I_a,\J_b}$ is separated by
$\delta_1(\set{z,u'})$, for some $u'\in\set{x,y,\Max}$.
\\
Since 
\[
 \dist(a,\Min^{\A}) + \delta_1(\set{z,u'})\quad \leq \quad
 \delta_1(\set{\Min,z}) + \delta_1(\set{z,u'}) \quad \leq \quad \tildelta(\set{\Min,u'}),
\]
it is straightforward to see that
$\struc{\I_a,\J_b}$, and also $\struc{\I,\J}$,
is separated by $\tildelta(\set{\Min,u'})$.
I.e., $\tildelta$ is a separator for $\struc{\I,\J}$.
\medskip\\
\textit{\underline{Case~2:} \ $\dist(\Max^{\B},\Min^{\B}) < \dist(a,\Min^{\A})$}.\\
Since 
\[ 
 \tildelta(\set{\Min,\Max}) \quad \geq \quad \delta_1(\set{\Min,z})
 \quad \geq \quad \dist(a,\Min^{\A}) \quad > \quad \dist(\Max^{\B},\Min^{\B}), 
\]
we know that
$\struc{\I,\J}$ is separated by $\tildelta(\set{\Min,\Max})$.
I.e., $\tildelta$ is a separator for $\struc{\I,\J}$.
\\
This completes the proof of Claim~\ref{claim:3}.~\qed
\end{proof}%
\medskip \\
%
By symmetry we also obtain the following
\\
\begin{claim}~\label{claim:4}
If \,$\delta_1(\set{z,\Max})\geq \dist(\Max^{\A},a)$,\, then
$\tildelta$ is a separator for $\struc{\I,\J}$.~\fertig 
\end{claim}%
%
In a similar way, we can also show the following
\\
%
\begin{claim}~\label{claim:5}
If \,$\delta_1(\set{x,z})\geq |\dist(\alpha(x),a)|$,\, then
$\tildelta$ is a separator for $\struc{\I,\J}$.~\fertig 
\end{claim}%
\begin{proof}
We distinguish between three cases.
An illustration is given in Figure~\ref{fig:Claim5}.
\begin{figure}
\begin{center}
  \includegraphics[width=12cm]{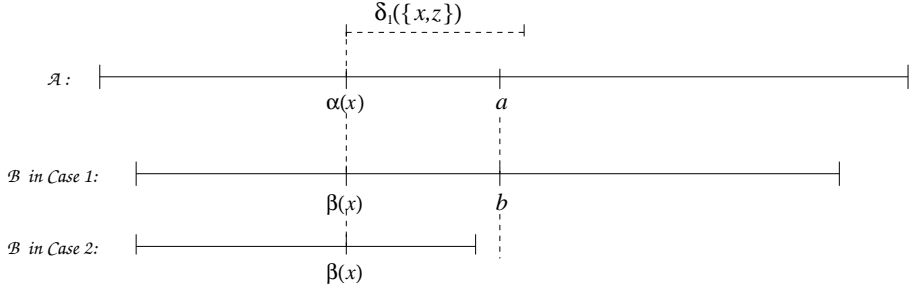}
\end{center}
\caption{Situation in Claim 5 (for the special case that $\alpha(x)\leq a$).}\label{fig:Claim5}
\end{figure}
\medskip\\
\textit{\underline{Case~1:} \ There is a $b\in\U^{\B}$ such that
$\dist(\beta(x),b) = \dist(\alpha(x),a)$.}
\\
Obviously, $\struc{\I_a,\J_b}$ is not separated by $\delta_1(\set{x,z})$.
However, we know that $\struc{\I_a,\J_b}$ is separated by
$\delta_1(\set{z,u'})$, for some $u'\in\set{\Min,y,\Max}$.
\\
Since 
\[
 |\dist(\alpha(x),a)| + \delta_1(\set{z,u'}) \quad \leq \quad
 \delta_1(\set{x,z}) + \delta_1(\set{z,u'}) \quad \leq \quad \tildelta(\set{x,u'}),
\]
it is straightforward to see that
$\struc{\I_a,\J_b}$, and also $\struc{\I,\J}$,
is separated by $\tildelta(\set{x,u'})$.
I.e., $\tildelta$ is a separator for $\struc{\I,\J}$.
\medskip\\
\textit{\underline{Case~2:} \ $\alpha(x)\leq a$ \ \ and \ \  
   $\dist(\Max^{\B},\beta(x))< \dist(a,\alpha(x))$}.
\\
Since 
\[
 \tildelta(\set{x,\Max})\quad \geq \quad \delta_1(\set{x,z}) 
 \quad \geq \quad \dist(a,\alpha(x)) 
 \quad > \quad \dist(\Max^{\B},\beta(x)), 
\]
we know that
$\struc{\I,\J}$ is separated by $\tildelta(\set{x,\Max})$.
I.e., $\tildelta$ is a separator for $\struc{\I,\J}$.
\medskip\\
\textit{\underline{Case~3:} \ $a<\alpha(x)$ \ \ and \ \ 
   $\dist(\beta(x),\Min^{\B})< \dist(\alpha(x),a)$}.
\\
This case is analogous to Case~2.
\smallskip\\
Now the proof of Claim~\ref{claim:5} is complete, 
because one of the three cases above must apply.~\qed 
\end{proof}%
\medskip\\
%
By symmetry we also obtain the following
\\
%
\begin{claim}~\label{claim:6}
If \,$\delta_1(\set{y,z})\geq |\dist(\alpha(y),a)|$,\, then
$\tildelta$ is a separator for $\struc{\I,\J}$.~\fertig 
\end{claim}%
%
Finally, we show the following
\\
%
\begin{claim}~\label{claim:7}
If none of the assumptions of the Claims~\ref{claim:1}--\ref{claim:6}
is satisfied, then $\tildelta$ is a separator for $\struc{\I,\J}$.~\fertig
\end{claim}%
\begin{proof}
We assume w.l.o.g.\ that 
\ $\Min^{\A}\leq \alpha(x) \leq \alpha(y) \leq \Max^{\A}$.
\\
Since Claim~\ref{claim:1} does not apply, we know that also
\ $\Min^{\B}\leq \beta(x) \leq \beta(y) \leq \Max^{\B}$.
\\
Since Claims~\ref{claim:2}--\ref{claim:6} do not apply, we
furthermore know that
\begin{enumerate}[1. \ ]
\item
 $|\dist(\alpha(u'),a)| \ > \ \delta_1(\set{u',z})$,\quad for all $u'\in\set{\Min,\Max,x,y}$,\quad and
\item
 $\Min^{\A} < a < \alpha(x)$ \quad or \quad 
 $\alpha(x) < a < \alpha(y)$ \quad or \quad 
 $\alpha(y) < a < \Max^{\A}$.
\end{enumerate}%
We distinguish between different cases, depending on the 
particular interval that $a$ belongs to.
An illustration is given in Figure~\ref{fig:Claim7-1}.
\begin{figure}
\begin{center}
  \includegraphics[width=12cm]{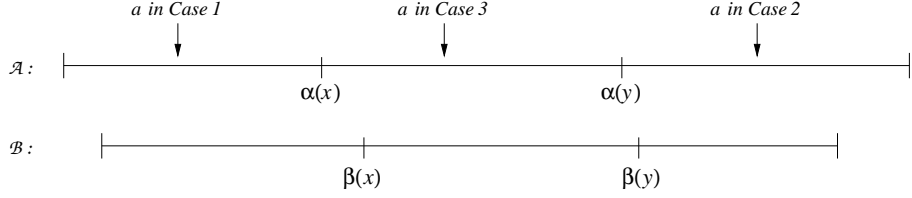}
\end{center}
\caption{Situation at the beginning of Claim 7.}\label{fig:Claim7-1}
\end{figure}
\medskip\\
\textit{\underline{Case~1:} \  $\Min^{\A} < a < \alpha(x)$.}
\medskip\\
 \textit{\underline{Case~1.1:} \ $\dist(\beta(x),\Min^{\B})\leq \delta_1(\set{\Min,z})$.}\\
 By definition of $\tildelta$ we have 
 $\dist(\beta(x),\Min^{\B}) \leq \tildelta(\set{\Min,x})$.
 Since 
 \[
 \dist(\alpha(x),\Min^{\A}) \quad  > \quad \dist(a,\Min^{\A}) \quad > \quad \delta_1(\set{\Min,z}) 
  \quad \geq \quad \dist(\beta(x),\Min^{\B}),
  \]
  we therefore know that 
  $\struc{\I,\J}$ is separated by $\tildelta(\set{\Min,z})$. 
 I.e., $\tildelta$ is a separator for $\struc{\I,\J}$.
\medskip\\
 \textit{\underline{Case~1.2:} \ $\dist(\beta(x),\Min^{\B}) > \delta_1(\set{\Min,z})$.}\\
 In this case we can choose $b\leq \beta(x)$ such that
 $\dist(b,\Min^{\B}) = \delta_1(\set{\Min,z})+1$.
 An illustration is given in Figure~\ref{fig:Claim7-1.2}.
\begin{figure}
\begin{center}
  \includegraphics[width=12cm]{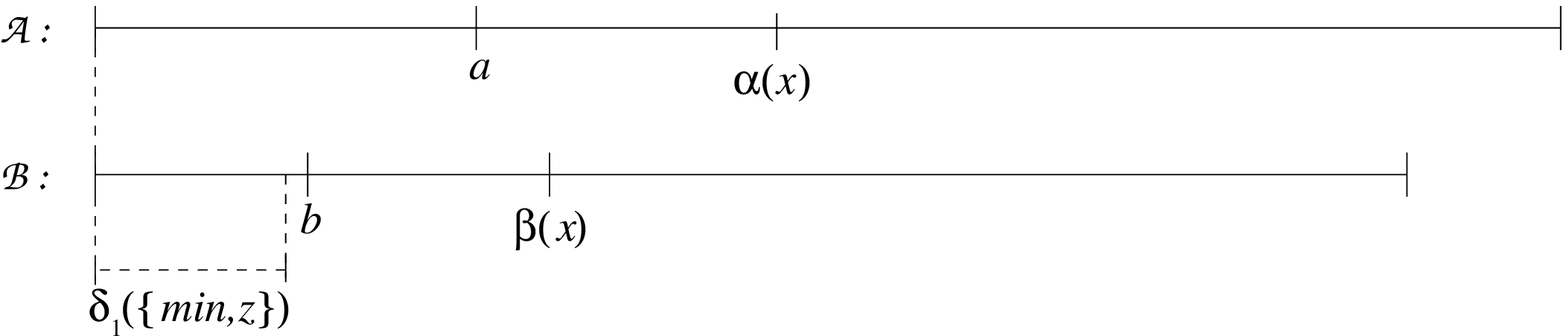}
\end{center}
\caption{Situation in Case 1.2 of Claim 7.}\label{fig:Claim7-1.2}
\end{figure}
 Then, $\struc{\I_a,\J_b}$ is not separated by $\delta_1(\set{\Min,z})$.
 However, we know that $\struc{\I_a,\J_b}$ is separated by $\delta_1(\set{z,u'})$, 
 for some $u'\in\set{x,y,\Max}$.
 From the assumptions of Claim~\ref{claim:7} we know that 
 $\dist(\alpha(u'),a)>\delta_1(\set{z,u'})$.
 Hence we must have that $\dist(\beta(u'),b)\leq \delta_1(\set{z,u'})$.
 Therefore, 
 \begin{eqnarray*} 
 \dist(\beta(u'),\Min^{\B})  & = & \dist(\beta(u'),b) + \dist(b,\Min^{\B}) \\
 & \leq &  \delta_1(\set{z,u'}) + \delta_1(\set{\Min,z}) + 1  \\
 & \leq & \tildelta(\set{\Min,u'}).
 \end{eqnarray*}
 Since 
 \[
  \dist(\alpha(u'),\Min^{\A}) \quad = \quad
  \dist(\alpha(u'),a) + \dist(a,\Min^{\A})  \quad > \quad \dist(\beta(u'),\Min^{\B}),
 \] 
 we hence obtain that $\tildelta$ is a separator for $\struc{\I,\J}$.
\medskip\\
\textit{\underline{Case~2:} \  $\alpha(y) < a < \Max^{\A}$.}\\
This case is analogous to Case~1.
\medskip\\
\textit{\underline{Case~3:} \  $\alpha(x) < a < \alpha(y)$.}
\medskip\\
 \textit{\underline{Case~3.1:} \ $\dist(\beta(y),\beta(x))\leq \delta_1(\set{x,z})$.}\\
 By definition of $\tildelta$ we have 
 $\dist(\beta(y),\beta(x)) \leq \tildelta(\set{x,y})$.
 Since 
 \[
 \dist(\alpha(y),\alpha(x)) \quad > \quad \dist(a,\alpha(x)) \quad > \quad \delta_1(\set{x,z}) 
 \quad \geq \quad \dist(\beta(y),\beta(x)),
  \]
we therefore know that 
 $\struc{\I,\J}$ is separated by $\tildelta(\set{x,y})$. 
 I.e., $\tildelta$ is a separator for $\struc{\I,\J}$.
\medskip\\
 \textit{\underline{Case~3.2:} \ $\dist(\beta(y),\beta(x)) > \delta_1(\set{x,z})$.}\\
 In this case we can 
 choose $b$ with $\beta(x)< b \leq \beta(y)$ such that
 $\dist(b,\beta(x)) = \delta_1(\set{x,z})+1$.
 An illustration is given in Figure~\ref{fig:Claim7-3.2}.
\begin{figure}
\begin{center}
  \includegraphics[width=12cm]{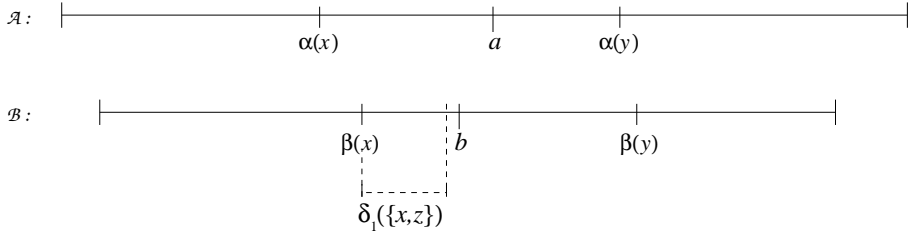}
\end{center}
\caption{Situation in Case 3.2 of Claim 7.}\label{fig:Claim7-3.2}
\end{figure}
 Then, $\struc{\I_a,\J_b}$ is not separated by $\delta_1(\set{x,z})$.
 However, we know that $\struc{\I_a,\J_b}$ is separated by $\delta_1(\set{z,u'})$, 
 for some $u'\in\set{\Min,y,\Max}$.
 From the assumptions of Claim~\ref{claim:7} we know 
 that $|\dist(a,\alpha(u'))|>\delta_1(\set{z,u'})$.
 Hence we must have that $|\dist(b,\beta(u'))|\leq \delta_1(\set{z,u'})$.
 We now distinguish between the cases 
 where $u'$ can be chosen from $\set{y,\Max}$, on the one hand, and
 where $u'$ must be chosen as $\Min$, on the other hand.
 \medskip\\
 \textit{\underline{Case~3.2.1:} \ $u' \in \set{y,\Max}$.}\\
 In this case,
 \begin{eqnarray*}
 \dist(\beta(u'),\beta(x)) & = & \dist(\beta(u'),b) + \dist(b,\beta(x)) \\
 & \leq & \delta_1(\set{z,u'}) + \delta_1(\set{x,z}) + 1   \\
 & \leq & \tildelta(\set{x,u'}).
 \end{eqnarray*}
 Since 
 \[
 \dist(\alpha(u'),\alpha(x)) \quad = \quad
 \dist(\alpha(u'),a) + \dist(a,\alpha(x)) \quad > \quad \dist(\beta(u'),\beta(x)),
 \] 
 we hence obtain that $\tildelta$ is a separator for $\struc{\I,\J}$.
 \medskip\\
 \textit{\underline{Case~3.2.2:} \ $u'\not\in \set{y,\Max}$.} \\
 In this case, $\struc{\I_a,\J_b}$ is separated by $\delta_1(\set{\Min,z})$, 
 and we may assume that it is neither separated by $\delta_1(\set{z,y})$ nor
 by $\delta_1(\set{z,\Max})$ nor 
 by $\delta_1(\set{x,z})$. 
 In particular, we must have that
 \[
 \dist(\beta(y),b) \quad \geq \quad \delta_1(\set{z,y}) + 1.
 \]
 Therefore, for every $b'$ with 
 \[
   b \quad \leq \quad  b' \quad < \quad \beta(y)-\delta_1(\set{z,y}),
 \] 
 the
 following is true: $\struc{\I_a,\J_{b'}}$ is neither separated by
 $\delta_1(\set{x,z})$ nor by $\delta_1(\set{z,y})$, but, consequently,
 by $\delta_1(\set{\Min,z})$ or by $\delta_1(\set{z,\Max})$.
 Let $b_1$ be the largest such $b'$ for which $\struc{\I_a,\J_{b'}}$ is 
 separated by $\delta_1(\set{\Min,z})$. 
 In particular, 
 \[ 
   \dist(b_1,\Min^{\B}) \quad \leq \quad \delta_1(\set{\Min,z}).
 \]
 An illustration is given in Figure~\ref{fig:Claim7-3.2.2}.
\begin{figure}
\begin{center}
  \includegraphics[width=12cm]{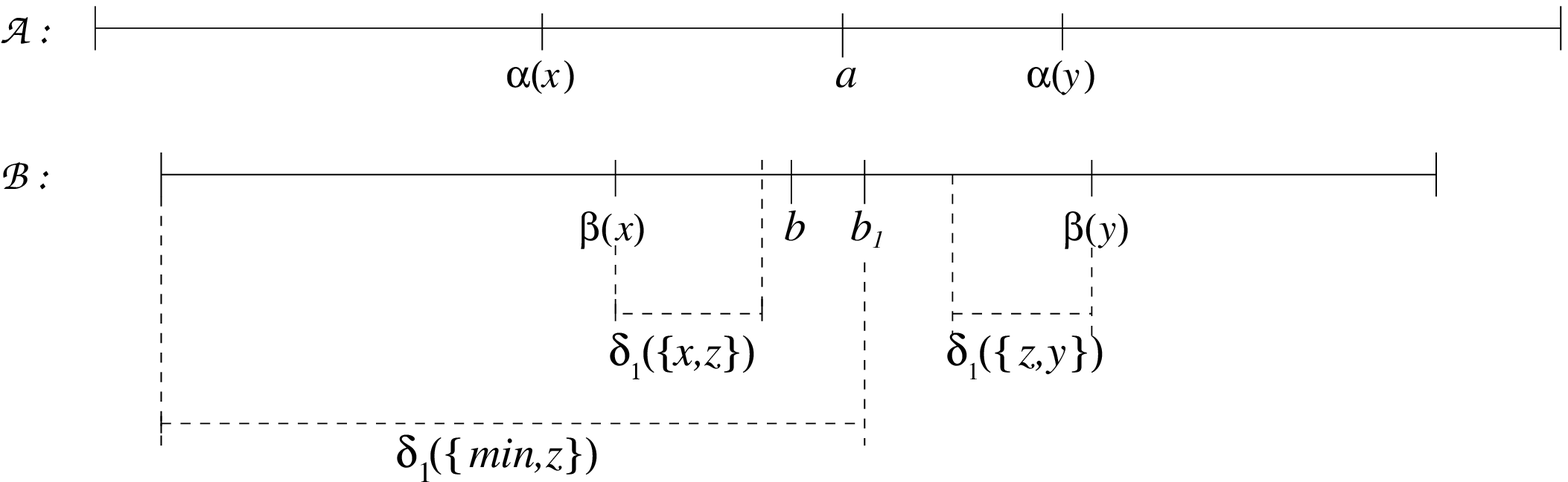}
\end{center}
\caption{Situation in Case 3.2.2 of Claim 7.}\label{fig:Claim7-3.2.2}
\end{figure}
 \medskip\\
 \textit{\underline{Case~3.2.2.1:} \ $\dist(\beta(y),b_1{+}1)\leq\delta_1({\set{z,y}})$.}
 \\
 In this case we know that
 \[
  \dist(\beta(y),\Min^{\B}) \quad \leq \quad \delta_1(\set{z,y}) + 1 + \delta_1(\set{\Min,z}) 
  \quad \leq \quad \tildelta(\set{\Min,y}).
 \]
 Furthermore, 
 \begin{eqnarray*}
 \dist(\alpha(y),\Min^{\A}) & = &
 \dist(\alpha(y),a) + \dist(a,\Min^{\A}) \\ 
 & \geq &
 \delta_1(\set{z,y}) + 1 + \delta_1(\set{\Min,z}) + 1.
 \end{eqnarray*}
 Therefore, 
 \[ 
   \dist(\alpha(y),\Min^{\A}) \quad \neq \quad \dist(\beta(y),\Min^{\B}),
 \] 
 and $\struc{\I,\J}$ is
 separated by $\tildelta(\set{\Min,y})$. I.e., 
 $\tildelta$ is a separator for $\struc{\I,\J}$.%
 \medskip\\
\textit{\underline{Case~3.2.2.2:} \ $\dist(\beta(y),b_1{+}1) > \delta_1({\set{z,y}})$.}
 \\ 
 In this case we know (by the maximal choice of $b_1$) that
 $\struc{\I_a,\J_{b_1+1}}$ must be separated by $\delta_1(\set{z,\Max})$.
 In particular, \ $\dist(\Max^{\B},b_1{+}1)\leq \delta_1({z,\Max})$.
 Therefore, 
 \[
 \dist(\Max^{\B},\Min^{\B}) \quad \leq \quad
  \delta_1(\set{z,\Max}) + 1 + \delta_1(\set{\Min,z}) \quad \leq \quad
  \tildelta(\set{\Min,\Max}).
  \]
 Furthermore, 
 \begin{eqnarray*}
 \dist(\Max^{\A},\Min^{\A}) & \geq &
  \dist(\Max^{\A},a) + \dist(a,\Min^{\A}) \\
 & \geq & 
 \delta_1(\set{z,\Max}) +1 + \delta_1(\set{\Min,z}) +1.
 \end{eqnarray*}
 Therefore, 
 \[
  \dist(\Max^{\A},\Min^{\A}) \quad \neq \quad
  \dist(\Max^{\B},\Min^{\B}),
 \] 
 and $\struc{\I,\J}$ is separated by
 $\tildelta(\set{\Min,\Max})$. 
 I.e., $\tildelta$ is a separator for $\struc{\I,\J}$. 
 \medskip\\
 We now have shown that $\tildelta$ is a
 separator for $\struc{\I,\J}$, if Case~3 applies.
\\
Together with the Cases~1 and 2 we therefore obtain that
the proof of Claim~\ref{claim:7} is complete.\\ \mbox{}\qed
\end{proof}%
\medskip\\
%
Since at least one of the Claims~\ref{claim:1}--\ref{claim:7}
must apply, the proof of Lemma~\ref{lemma:separator-one} finally is complete.~\qed
\end{proof}
%
%
\\
\parno
%
%
\begin{proofc}{of part \emph{(\ref{key-prop:one})} of Lemma~\ref{lemma:key-prop}}\mbox{}\\
Let $v$ be a node of $\T$ that has exactly one child $v_1$. 
Let $\delta$ be a minimal separator for $\Il(v)$, and let 
$\delta_1$ be a minimal separator for $\struc{A_1,B_1}\deff \Il(v_1)$.
Our aim is to show that \ 
$w(\delta) \ \leq \ w(\delta_1) + 2$.
\par
From Definition~\ref{definition:extended_syntax_tree} we know that either \,$\Fl(v)=\nicht$\, or
\,$\Fl(v)=\textsf{\upshape Q}u$,\, for some $\textsf{\upshape Q}\in\set{\exists,\forall}$ and
$u\in\set{x,y,z}$.
\medskip\\
\textit{Case~1:} $\Fl(v)=\nicht$\\
In this case we know from Definition~\ref{definition:extended_syntax_tree} 
that $\Il(v)= \struc{B_1,A_1}$. Therefore, $\delta_1$ also is a (minimal) separator for
$\Il(v)$. In particular, 
\ $w(\delta) = w(\delta_1) \leq \ w(\delta_1)+ 2$.
\medskip\\
\textit{Case~2:} $\Fl(v)=\textsf{\upshape Q}u$\\
In this case let $\tildelta$ be the \emph{separator for $\Il(v)$} defined in
Lemma~\ref{lemma:separator-one}. Since $\delta$ is a \emph{minimal} separator for $\Il(v)$,
it suffices to show that
\ $w(\tildelta)\ \leq \ w(\delta_1) + 2$.
\par
Let $u',u''$ be chosen such that $\set{x,y,z}=\set{u,u',u''}$.
Using Definition~\ref{definition:weight} and the particular choice of $\tildelta$, it
is straightforward to see that 
{
\begin{equation}\label{Proof_key-prop:one:item:1}
 c(\tildelta) \quad = \quad \tildelta(\set{u',u''}) \quad \leq \quad  c(\delta_1) + 1
\end{equation}%
}
and that
{
\begin{equation}\label{Proof_key-prop:one:item:2}
 \tildelta(\set{\Min,\Max})\ \ \leq \ \ b(\delta_1) +1\,.
\end{equation}%
}
Furthermore, for arbitrary $\tilde{u},\tilde{u}' \in \set{x,y,z}$ we have 
{
\begin{equation}\label{Proof_key-prop:one:item:3}
\tildelta(\set{\Min,\tilde{u}}) + \tildelta(\set{\tilde{u}',\Max})
\ \ \leq \ \  
b(\delta_1) + 2 c(\delta_1) + 2\,,
\end{equation}
}
which can be seen as follows:
If \ $\tilde{u}= u$ \ or \ $\tilde{u}'=u$, \ then \
$\tildelta(\set{\Min,\tilde{u}}) = 0$ \ or \ $\tildelta(\set{\tilde{u}',\Max})= 0$.
\\
Consequently, 
\[
 \tildelta(\set{\Min,\tilde{u}})+\tildelta(\set{\tilde{u}',\Max}) \quad \leq \quad
  \MAX \set{b(\delta_1),\ b(\delta_1)+c(\delta_1)+1}.
\]
If $\tilde{u}$ and $\tilde{u}'$ both are different from $u$, then
\begin{eqnarray*}
  \tildelta(\set{\Min,\tilde{u}}) & = &
  \MAX \; \big\{\ 
      \delta_1(\set{\Min,\tilde{u}}), \ \ 
      \delta_1(\set{\Min,u}) + \delta_1(\set{u,\tilde{u}}) +1
    \ \big\}
\end{eqnarray*}%
and
\begin{eqnarray*}
  \tildelta(\set{\tilde{u}',\Max}) & = &
  \MAX \; \big\{\ 
      \delta_1(\set{\tilde{u}',\Max}),\ \
      \delta_1(\set{u,\Max}) + \delta_1(\set{u,\tilde{u}'}) +1
    \ \big\}\,.
\end{eqnarray*}
Therefore,
\ $\tildelta(\set{\Min,\tilde{u}}) + \tildelta(\set{\tilde{u}',\Max}) \ \leq$
\[
  \MAX \; 
  \left\{
      \begin{array}{l}
        \delta_1(\set{\Min,\tilde{u}}) +
        \delta_1(\set{\tilde{u}',\Max}),
        \\[1ex]
        \delta_1(\set{\Min,\tilde{u}}) +
        \delta_1(\set{u,\Max}) + \delta_1(\set{u,\tilde{u}'}) +1,
        \\[1ex] 
        \delta_1(\set{\Min,u}) + \delta_1(\set{u,\tilde{u}}) +1 +
        \delta_1(\set{\tilde{u}',\Max}),
        \\[1ex]
        \delta_1(\set{\Min,u}) + \delta_1(\set{u,\tilde{u}}) +1 +
           \delta_1(\set{u,\Max}) + \delta_1(\set{u,\tilde{u}'}) +1
      \end{array}
  \right\}
\]
which, in turn, is less than or equal to
\[
 \MAX \; \big\{\ b(\delta_1),\ \
                 b(\delta_1) + c(\delta_1) + 1,\ \
                 b(\delta_1) + 2 c(\delta_1) + 2 
      \ \big\}\,.
\]
I.e., we have shown that (\ref{Proof_key-prop:one:item:3}) is valid.
\\
From (\ref{Proof_key-prop:one:item:2}) and (\ref{Proof_key-prop:one:item:3}) we obtain 
\begin{equation}\label{Proof_key-prop:one:item:4}
 b(\tildelta)\quad \leq \quad  b(\delta_1) + 2 c(\delta_1) + 2\,.
\end{equation}%
From (\ref{Proof_key-prop:one:item:1}) and (\ref{Proof_key-prop:one:item:4}) we conclude that
\begin{eqnarray*}
  w(\tildelta)^2
&  \ = \ 
& c(\tildelta)^2 + b(\tildelta) 
\\
&  \leq
& \big( c(\delta_1) + 1 \big)^2 + b(\delta_1) + 2c(\delta_1)+ 2
\\
&  =
& c(\delta_1)^2 + b(\delta_1) + 4 c(\delta_1) + 3
\\
&  \leq
& w(\delta_1)^2 + 4 w(\delta_1) + 4 
\\
& =
& \big( w(\delta_1) + 2 \big)^2\,.
\end{eqnarray*}
Therefore, \ $w(\tildelta) \ \leq \ w(\delta_1) + 2$.
\\
This completes the proof of  
part \emph{(\ref{key-prop:one})} of Lemma~\ref{lemma:key-prop}.
\qed
\end{proofc}
%
%
%
\begin{remark}\label{remark:choice_of_weight}
From the above proof it becomes clear, why 
Definition~\ref{definition:weight} fixes the \emph{weight} of
a separator by using the $\sqrt{\ \,}$-function.
Let us consider the, at first glance, more straightforward weight function
$\hat{w}(\delta)\deff \MAX \set {c(\delta),\ b(\delta)}$. 
In the proof of part \emph{(\ref{key-prop:one})} of Lemma~\ref{lemma:key-prop}
we then obtain from the items (\ref{Proof_key-prop:one:item:1}) and (\ref{Proof_key-prop:one:item:4})
that $\hat{w}(\delta) \leq 2c(\delta_1)+ b(\delta_1) + 2 \leq 3\hat{w}(\delta_1)+2$.
Therefore, a modified version of Lemma~\ref{lemma:weighted_trees},
where item (c) is replaced by the condition ``\emph{If $v$ has exactly one child $v_1$, then
  $w(v) \ \ \leq \ \ 3w(v_1) \ + \ 2$}'', leads to a (much weaker) bound of the form 
$\size{\T}\geq c\cdot \Log(w(v))$. This, in turn, leads to a weaker version of
Theorem~\ref{theorem:lower_bound}, stating that $\size{\psi}\geq c\cdot\Log\big(\hat{w}(\delta)\big)$.
However, this bound can already be proven by a conventional Ehrenfeucht-Fra\"\i{}ss\'{e} game and
does not only apply for $\FO^3(<)$-formulas but for $\FO(<)$-formulas in general and therefore
is of no use for comparing the succinctness of $\FO^3$ and $\FO$.
~\fertig
\end{remark}%
\section{$\bs{\FO^2}$ vs.\ $\bs{\FO^3}$}\label{section:FO2vsFO3}%
%
%
%
As a first application, Theorem~\ref{theorem:lower_bound} allows us to translate every
$\FO^3$-sentence $\psi$ into an $\FO^2$-sentence $\chi$ that is
equivalent to $\psi$ on 
linear orders and that has size polynomial in the size of $\psi$.
To show this, we use the following easy lemmas.
\begin{lemma}\label{lemma:FOkleiner}
Let $\varphi$ be an $\FO(<,\allowbreak \Succ,\allowbreak \Min,\allowbreak \Max)$-sentence,
and let $d$ be the quantifier depth of $\varphi$. 
For all $N\geq 2^{d+1}$,
\,$\A_N\models \varphi$\, if, and only if, \,$\A_{2^{d+1}}\models \varphi$.~\fertig
\end{lemma}%
A proof can be found, e.g., in the textbook \cite{EbbinghausFlum}.
\begin{lemma}\label{lemma:FOzwei-length-def}
For all $\ell\in\NN$ there are
$\FO^2(<)$-sentences $\chi_{\ell}$ and
$\chi_{\geq \ell}$ of size $\bigO(\ell)$ such that, for all $N\in\NN$, \
$\A_N\models \chi_{\ell}$ \,(respectively, $\chi_{\geq \ell}$) 
\;iff \,$N=\ell$ 
\,(respectively, $N\geq \ell$).~\fertig 
\end{lemma}%
\begin{proof}
We choose 
\begin{eqnarray*} 
  \chi'_{\geq 0}(x) & \deff & (x=x),
\end{eqnarray*}
and, 
for all $\ell \geq 0$, 
\begin{eqnarray*}
 \chi'_{\geq\ell+1}(x) & \deff &
 \exists\; y\ (y<x) \ \und \ \chi_{\geq\ell}(y).
\end{eqnarray*}
Obviously, for all $N\in\NN$ and all $a\in\set{0,\twodots,N}$, we have
$\A_N\models\chi'_{\geq \ell}(a)$ iff $a\geq\ell$. 
Therefore, for every $\ell\in\NN$, we can choose
\ $\chi_{\geq\ell}\deff \exists\, x\,\chi'_{\geq\ell}(x)$ \ and
\ $\chi_{\ell}\deff \chi_{\geq\ell}\und\nicht\chi_{\geq \ell+1}$. 
\qed
\end{proof}%
\begin{theorem}\label{theorem:FO2vsFO3}\mbox{}\\
On linear orders, $\FO^3(<,\Succ,\Min,\Max)$-sentences are
$\bigO(m^4)$-succinct in
$\FO^2(<)$-sen\-ten\-ces.\fertig 
\end{theorem}%
\begin{proof}
Let $\psi$ be an $\FO^3(<,\Succ,\Min,\Max)$-sentence.
Our aim is to find an $\FO^2(<)$-sentence $\chi$ of size $\bigO(\size{\psi}^4)$ 
such that, for all $N\in\NN$, $\A_N\models\chi$ iff $\A_N\models\psi$.
\par
If $\psi$ is satisfied by \emph{all} linear orders or by \emph{no} linear order,
$\chi$ can be chosen in a straightforward way. In all other cases we know
from Lemma~\ref{lemma:FOkleiner} that there exists a $D\in\NN$ such that
either 
\begin{enumerate}[(1.)\ ]
\item
$\A_{D}\models\psi$\, and, for all $N>D$, \,$\A_N\not\models\psi$,
\quad or
\item
$\A_{D}\not\models\psi$\, and, for all $N>D$, \,$\A_N\models\psi$.
\end{enumerate}
In particular, $\psi$ is an $\FO^3$-sentence that distinguishes 
between the linear orders $\A_D$ and $\A_{D+1}$. From Corollary~\ref{cor:lower_bound}
we therefore know that $\size{\psi}\geq\einhalb \sqrt{D}$.
\par
We next construct an $\FO^2$-sentence $\chi$ equivalent to $\psi$:
Let $\chi_{\ell}$ and $\chi_{\geq\ell}$ be the $\FO^2(<)$-sentences
from Lemma~\ref{lemma:FOzwei-length-def}.
Let $\chi'$ be the disjunction of the sentences $\chi_{\ell}$ for
all those $\ell\leq D$ with $\A_{\ell}\models\psi$. 
Finally, if $\A_{D+1}\models\psi$, then choose
$\chi\deff \chi'\oder\chi_{\geq D+1}$; otherwise choose $\chi\deff \chi'$.
Obviously, 
$\chi$ is an $\FO^2(<)$-sentence equivalent to $\psi$, and
\[  
  \size{\chi} \quad =  \quad \bigO\big(\sum_{\ell = 0}^{D+1} \ell \big)
  \quad = \quad \bigO\big(D^2\big)
  \quad = \quad \bigO(\size{\psi}^4).
\]
This completes the proof of Theorem~\ref{theorem:FO2vsFO3}.
\qed
\end{proof}%
\medskip
\section{$\bs{\FO^3}$ vs.\ $\bs{\FO}$}\label{section:FOvsFO3}%
%
%
Using Theorem~\ref{theorem:lower_bound}, we will show in this section 
that there
is an exponential succinctness gap between $\FO$ and $\FO^3$ on linear orders.
\begin{lemma}\label{lemma:FO_to_FO3}
For every $\FO(<,\Succ,\Min,\Max)$-sentence $\varphi$ there is an $\FO^2(<)$-sentence
$\psi$ of size $\size{\psi} \leq 2^{\bigO(\size{\varphi})}$ which is
equivalent to $\varphi$ on the class of linear orders.
\fertig
\end{lemma}%
\begin{proof}
Let $\varphi$ be an $\FO(<,\allowbreak \Succ,\allowbreak \Min,\allowbreak\Max)$-sentence,
and let $d$ be the 
quantifier depth of $\varphi$. In particular, $\size{\varphi}\geq d$.
From Lemma~\ref{lemma:FOkleiner} we know 
that, for all $N\geq 2^{d+1}$,
$\A_N\models \varphi$ if, and only if, $A_{2^{d+1}}\models \varphi$.
\par
We use, for every $\ell\in\NN$, the sentences $\chi_{\ell}$ and $\chi_{\geq\ell}$ of 
Lemma~\ref{lemma:FOzwei-length-def}. 
Let $\psi'$ be the disjunction of the sentences $\chi_{\ell}$ for
all $\ell<2^{d+1}$ such that $\A_{\ell}\models\varphi$. 
\\
Finally, if $\A_{2^{d+1}}\models\varphi$, then choose
$\psi\deff \psi'\oder\chi_{\geq 2^{d+1}}$; otherwise choose $\psi\deff \psi'$.
Obviously, 
$\psi$ is an $\FO^2(<)$-sentence equivalent to $\varphi$, and
\[
 \size{\psi} \quad = \quad
 \bigO\big(\sum_{\ell = 0}^{2^{d+1}} \ell \big) \quad = \quad
 \bigO\big(2^{2(d+1)}\big) \quad = \quad 
 2^{\bigO(\size{\varphi})}.
\]
\mbox{ }\qed
\end{proof}
\begin{lemma}\label{lemma:FO-succinctness}
For all $m\in\NN$ there are 
$\FO^4(<)$-sentences $\varphi_m$ and sets $A_m$ and $B_m$ of interpretations, 
such that $A_m\models \varphi_m$, $B_m\models \nicht\varphi_m$,
$\size{\varphi_m} = \bigO(m)$, and every $\FO^3(<,\Succ,\Min,\Max)$-sentence $\psi_m$
equivalent to $\varphi_m$ has size 
$\size{\psi_m}\geq 2^{\frac{1}{2}m-1}$.
\fertig
\end{lemma}%
\begin{proof}
For every $N\in \NN$ 
let $\alpha_N : \set{\Min,x,y,z,\Max} \rightarrow \set{0,\twodots,N}$ be the 
assignment with $\alpha_N(x) = \alpha_N(\Min) = 0$ \ and \ $\alpha_N(y)=\alpha_N(z)=\alpha(\Max) = N$.
\par
For every $m\in \NN$ we choose $A_m \deff \bigset{(\A_{2^m},\alpha_{2^m})}$
and $B_m \deff \bigset{(\A_{2^m+1},\alpha_{2^m+1})}$.
\medskip\\
\emph{Step~1: Choice of $\varphi_m$.}\\
Inductively we define $\FO^4(<)$-formulas $\varphi'_m(x,y)$ expressing
that $|\dist(x,y)| = 2^m$ via 
\begin{eqnarray*}
 \varphi'_{m}(x,y) & := &
   \exists\, z\ \forall\, u\ \ (u=x \,\oder\, u=y)\;\impl\; \varphi'_{m-1}(z,u)
\end{eqnarray*}
(and $\varphi'_0(x,y)$ chosen appropriately).
\\
It is straightforward to see that $\size{\varphi'_m} = \bigO(m)$ and
that $\varphi'_m(x,y)$ expresses that $|\dist(x,y)|= 2^m$.
\\
Therefore, 
\begin{eqnarray*}
 \varphi_m & \deff & \exists\, x\ \exists\, y\ \ \varphi'_m(x,y) \ \und \
 \nicht\,\exists\, z \ (z{<}x \,\oder\, y{<}z)
\end{eqnarray*}
is an $\FO^4(<)$-sentence
with the desired properties.
\medskip\\
\emph{Step~2: Size of equivalent $\FO^3$-sentences.}\\
For every $m\in\NN$ let $\psi_m$ be an $\FO^3(<,\Succ,\Min,\Max)$-sentence with
$A_m\models\psi_m$ and $B_m\models\nicht\psi_m$. 
From Corollary~\ref{cor:lower_bound} we conclude that 
\[
 \size{\psi_m}\quad \geq \quad \einhalb \sqrt{2^m}\quad =\quad 2^{\frac{1}{2} m -1}.
\]
This completes the proof of Lemma~\ref{lemma:FO-succinctness}.\mbox{}\qed
\end{proof}%
\medskip\\
From Lemma~\ref{lemma:FO_to_FO3} 
and~\ref{lemma:FO-succinctness} we directly obtain
\begin{theorem}\label{theorem:FO3vsFO}%
\mbox{}\\
On the class of linear orders,
\mbox{$\FO(<)$-}\allowbreak{}sentences are
$2^{\bigO(m)}$-succinct in $\FO^3(<)$-sentences, 
but already $\FO^4(<)$-sentences are not $2^{o(m)}$-succinct in
$\FO^3(<)$-sentences.\fertig 
\end{theorem}%
Note that the relation $\Succ$ and the constants $\Min$ and $\Max$ 
are easily definable in $\FO^3(<)$ and
could therefore be added in Theorem~\ref{theorem:FO3vsFO}.

\medskip

\section{$\FO$ vs.\ $\MSO$}\label{section:FOvsMSO}%
%
%
In this section we compare the succinctness of $\FO$ and the
$\FO$-expressible fragment of monadic second-order logic (for short: $\MSO$).
This section's main result is a non-elementary succintness gap between $\FO$ and
the $\FO$-expressible fragment of $\MSO$ on the class of linear orders 
(Theorem~\ref{theorem:FOvsMSO-UnaryAlphabet}).

The main idea for proving this succinctness gap is to encode natural numbers by strings in such
a way that there are extremely short $\MSO$-formulas for ``decoding'' these strings back into
numbers.
The method goes back to Stockmeyer and Meyer \cite{StockmeyerMeyer,sto74}; the particular
encoding used in the present paper has been introduced in \cite{FrickGrohe_JournalVersionOfLICS02}.
To formally state and prove this section's main result, we need some further notation:

We write $\MonSigma_1^1$ for the class of all $\MSO$-formulas that consist of
a prefix of existential set quantifiers, followed by a first-order formula.
By \,$\exists X\,\FO$\, we denote the fragment of $\MonSigma_1^1$ with only a
single existential set quantifier.
\par
Let $\Tower:\NN\rightarrow \NN$ be the function which maps every $h\in\NN$ to the
tower of 2s of height $h$. I.e., $\Tower(0)=1$ and, for every $h\in\NN$, $\Tower(h{+}1)=2^{\Tower(h)}$.
\par
We use the following notations of
\cite{FrickGrohe_JournalVersionOfLICS02}:
\\
For $h\geq 1$ let $\Sigma_h\deff
\big\{{0},{1},\Tag{1},\TagE{1},\twodots,\Tag{h},\TagE{h}\big\}$. 
The ``tags'' $\Tag{i}$ and $\TagE{i}$ represent single letters of the alphabet and are
just chosen to improve readability.
For every $n\geq 1$ let $L(n)$ be the length of the binary representation of the number 
$n{-}1$, i.e., $L(0)=0$, $L(1)=1$, and $L(n)=\abgerundet{\log(n{-}1)}+1$, for all $n\geq 2$.
By $\BIT(i,n)$ we denote the $i$-th bit of the binary representation of $n$, i.e.,
$\BIT(i,n)$ is 1 if $\abgerundet{\frac{n}{2^i}}$ is odd, and $\BIT(i,n)$ is $0$ 
otherwise.
\\
We encode every number $n\in\NN$ by a string $\mu_h(n)$ over the alphabet $\Sigma_h$,
where $\mu_h(n)$ is inductively defined as follows: \ 
\begin{eqnarray*}
 \mu_1(0) & \deff & \Tag{1}\ \TagE{1}\,, \quad \mbox{ and } \\
 \mu_1(n) & \deff &  
  \Tag{1}\ \;\BIT(0,n{-}1)\ \; \BIT(1,n{-}1) \ \cdots \ \BIT(L(n){-}1,n{-}1)\ \;\TagE{1}\,,
\end{eqnarray*}
for $n\geq 1$. For $h\geq 2$ we let 
\begin{eqnarray*}
 \mu_h(0) & \deff & \Tag{h}\ \TagE{h}\,, \quad \mbox{ and} \\
 \mu_h(n) & \deff &
     \Tag{h}\\
& &  \hphantom{\Tag{h}\,}\mu_{h-1}(0)\ \;\BIT(0,n{-}1) \\
& &  \hphantom{\Tag{h}\,}\mu_{h-1}(1)\ \;\BIT(1,n{-}1) \\
& &  \hphantom{\Tag{h}\, \mu_h} \vdots\\
& &  \hphantom{\Tag{h}\,} \mu_{h-1}(L(n){-}1)\ \;\BIT(L(n){-}1,n{-}1)\\
& &  \TagE{h}\,,
\end{eqnarray*}
for $n\geq 1$. Here, empty spaces and line breaks are just used to improve readability.
\par
For $h\in\NN$ let $H\deff \Tower(h)$. 
Let $\Sigma^{\bullet}_{h}\deff \Sigma_{h+1}\cup\set{\bullet}$, and let 
\begin{eqnarray*}
 v_h & \deff &
    \Tag{h{+}1}\ \  
     \mu_{h}(0)\, \bullet \ \ 
     \mu_{h}(1)\, \bullet \,
     \cdots \,
     \mu_h(H{-}1)\, \bullet\ \ 
   \TagE{h{+}1}\,.
\end{eqnarray*}
We consider the string-language
\,$(v_h)^+$, containing all strings that are the concatenation of one or more
copies of $v_h$. 
Let $w_h$ be the (unique) string in $(v_h)^+$
that consists of exactly $2^H$ copies of $v_h$.
\par
We write $\tau_h$ for the signature that consists of the symbol $<$ and a unary
relation symbol $P_{\sigma}$, for every $\sigma\in\Sigma_h^{\bullet}$.
Non-empty strings over $\Sigma_h^{\bullet}$ are represented by 
$\tau_h$-structures in the usual way (cf., e.g., \cite{EbbinghausFlum}).
We will shortly write \,$w\models\varphi$\, to indicate that the $\tau_h$-structure associated
with a $\Sigma_h^{\bullet}$-string $w$ satisfies a given $\tau_h$-sentence $\varphi$.
\par
The following lemma is our key tool for proving that there is a non-elementary 
succinctness gap between $\FO$ and the $\FO$-expressible fragment of $\MSO$.
%
\begin{lemma}\label{lemma:MSO-definability}
For every $h\in\NN$ there is an $\exists X\,\FO(\tau_h)$-sentence $\Phi_h$ 
of size $\bigO(h^2)$, such that the following
is true for all strings $w$ over the alphabet $\Sigma^{\bullet}_h$:
\ $w\models \Phi_h$ \ iff \allowbreak $w=w_h$.
\Fertig
\end{lemma}%
For proving Lemma~\ref{lemma:MSO-definability}
we need the following:
\begin{lemma}[{\cite[Lemma\;8]{FrickGrohe_JournalVersionOfLICS02}}]\label{lemma:equal-h}
  For all $h\in\NN$ there are $\FO(\tau_h)$-formulas \allowbreak 
  $\textit{equal}_{h}(x,y)$ of
  size\footnote{In
    \cite{FrickGrohe_JournalVersionOfLICS02}, an additional factor
    \,$\Log h$\, occurs because there a logarithmic cost measure is
    used for the formula size, whereas here we use a uniform measure.}  
  $\bigO(h)$ such that the following is true for all strings $w$ over 
  alphabet $\Sigma_h$, for all positions $a,b$ in $w$, and for all
  numbers $m,n\in\set{0,\twodots,\Tower(h)}$: If $a$ is the first
  position of a substring $u$ of $w$ that is isomorphic to $\mu_h(m)$
  and if $b$ is the first position of a substring $v$ of $w$ that is
  isomorphic to $\mu_h(n)$, then \ $w\models \textit{equal}_{h}(a,b)$ \ if, and
  only if, \,$m = n$.  \mbox{ }\Fertig
\end{lemma}%
Using the above lemma, it is an easy exercise to show
\begin{lemma}\label{lemma:inc-h}
  For every $h\in\NN$ there is an $\FO(\tau_h)$-formula $\textit{inc}_{h}(x,y)$ of
  size $\bigO(h)$ such that the following is true for all strings $w$ over
  alphabet $\Sigma_h$, for all positions $a,b$ in $w$, and for all
  numbers $m,n\in\set{0,\twodots,\Tower(h)}$: If $a$ is the first
  position of a substring $u$ of $w$ that is isomorphic to $\mu_h(m)$
  and if $b$ is the first position of a substring $v$ of $w$ that is
  isomorphic to $\mu_h(n)$, then \ $w\models \textit{inc}_{h}(a,b)$ \ if, and
  only if, \,$m{+}1 = n$.  \mbox{ }\Fertig
\end{lemma}%
We also need
\begin{lemma}\label{lemma:FO-def-of-vhPlus}
For every $h\in\NN$, the language $(v_h)^+$ is definable by an $\FO(\tau_h)$-sentence 
$\varphi_{(v_h)^+}$ of size $\bigO(h^2)$. I.e., for all strings $w$ over the alphabet 
$\Sigma^{\bullet}_h$ we have \,$w\models \varphi_{(v_h)^+}$ 
\,if, and only if,\, $w\in (v_h)^+$.\Fertig
\end{lemma}%
\begin{proof}
The proof proceeds in 2 steps:
\medskip\\
\underline{\emph{Step 1:}} \ 
Given $j\geq 1$, we say that a string $w$ over $\Sigma^{\bullet}_h$ satisfies the condition
$C(j)$ if, and only if, for every position $x$ (respectively, $y$) in $w$ that carries 
the letter $\Tag{j}$ (respectively, $\TagE{j}$) the following is true: 
There is a position $y$ to the right of $x$ that carries
the letter $\TagE{j}$ (respectively, a position $x$ to the left of $y$ that carries
the letter $\Tag{j}$), such that the substring $u$ of $w$ that starts at position $x$ 
and ends at position $y$ is of the form $\mu_j(n)$ for some 
$n\in\set{0,\twodots,\Tower(j){-}1}$.
\par
We will construct, for all $j\in\set{1,\twodots,h}$, 
$\FO(\tau_h)$-sentences
$\textit{ok}_j$ of size $\bigO(j)$ such that the following is true for all $j\leq h$ 
and all strings $w$ over $\Sigma^{\bullet}_h$ that satisfy the conditions $C(j')$ for all $j'<j$:
\[
  w \models \textit{ok}_j
  \quad\iff\quad
  w\, \mbox{ satisfies the condition } C(j).
\]
Simultaneously 
we will construct, for all $j\in\set{1,\twodots,h}$, 
$\FO(\tau_h)$-sentences
$\textit{max}_j(x)$ of size $\bigO(j)$ such that the following is true for all 
$j\leq h$, all strings
$w$ that satisfy the conditions $C(1),\twodots,C(j)$, and all positions $x$ in $w$:
\[
  w \models \textit{max}_j(x)
  \quad\iff \quad 
  \parbox[t]{6cm}{$x$ is the
     starting position of a substring of $w$ of the form $\mu_j(\Tower(j){-}1)$.}
\]
For the base case $j=1$ note that $\Tower(1){-}1 = 1$ and, by the definition of
$\mu_1(n)$,  
$\mu_1(0) = \Tag{1}\;\TagE{1}$ and $\mu_1(1)= \Tag{1}\;0\;\TagE{1}$. 
It is straightforward to write down a formula $\textit{ok}_1$ that expresses
the condition $C(1)$. Furthermore, $\textit{max}_1(x)$ states that the substring of
length 3 starting at position $x$ is of the form $\Tag{1}\;0\;\TagE{1}$. 
\par
For $j>1$ assume that the formula 
$\textit{max}_{j-1}$ has
already been constructed.
For the construction of the formula $\textit{ok}_j$ we 
assume that the underlying
string $w$ satisfies the conditions $C(1),\twodots,C(j{-}1)$.
\par
The formula $\textit{ok}_j$ states that whenever
$x$ (respectively, $y$) is a position in $w$ that carries 
the letter $\Tag{j}$ (respectively, $\TagE{j}$) the following is true: 
There is a position $y$ to the right of $x$ that carries
the letter $\TagE{j}$ (respectively, a position $x$ to the left of $y$ that carries
the letter $\Tag{j}$), such that the substring $u$ of $w$ that starts at position $x$ 
and ends at position $y$ is of the form $\mu_j(n)$ for some 
$n\in\set{0,\twodots,\Tower(j){-}1}$, i.e.,
\begin{enumerate}[\ 1.]
 \item
   the letters $\Tag{j}$ and $\TagE{j}$ only occur at the first and
   the last position of $u$,
 \item
   whenever a position $x'$ carries the letter $\TagE{j{-}1}$, position $x'{+}1$
   carries the letter $0$ or $1$, and position $x'{+}2$ carries the letter
   $\Tag{j{-}1}$ or the letter $\TagE{j}$,  
 \item
   either \,$u=\Tag{j}\;\TagE{j}$, or 
   the prefix of length 3 of $u$ is of the form 
   $\Tag{j}\;\Tag{j{-}1}\;\TagE{j{-}1}$, 
 \item
   whenever $x'$ and $y'$ are positions in $u$ carrying the letter $\Tag{j{-}1}$ such
   that $x'<y'$ and no position between $x'$ and $y'$ carries the letter $\Tag{j{-}1}$,
   the formula $\textit{inc}_{j-1}(x',y')$ from 
   Lemma~\ref{lemma:inc-h} is satisfied,
 \item
   if the rightmost position $x''$ in $u$ that carries the letter $\Tag{j{-}1}$ satisfies
   the formula $\textit{max}_{j-1}(x'')$, then there must be at least 
   one position $x'''$ 
   in $u$ that carries the letter $0$ such that $x'''{-}1$ carries the letter 
   $\TagE{j{-}1}$.
\end{enumerate}
Note that items 1.--4.\ ensure that $u$ is indeed of the form $\mu_j(n)$, for some
$n\in\NN$. Item~5 guarantees that $n\in\set{0,\twodots,\Tower(j){-}1}$ because of
the following:
recall from the definition of the string $\mu_j(n)$ that $\mu_j(n)$ involves the
(reverse) binary representation of the number $n{-}1$.
In particular, for $n\deff \Tower(j){-}1$, we need the (reverse) binary representation
of the number $\Tower(j){-}2$, which is of the form $011\cdots11$ and of length 
$\Tower(j{-}1)$, i.e., its highest bit has the number $\Tower(j{-}1){-}1$. 
\par
It is straightforward to see that the items 1.--5.\ and therefore also the formula
$\textit{ok}_j$ can be formalised by an $\FO(\tau_h)$-formula of size $\bigO(j)$, and that
this formula exactly expresses condition $C(j)$. 
\par
Furthermore, the formula $\textit{max}_j(x)$ assumes that $x$ is the starting
position of a substring $u$ of $w$ of the form $\mu_j(n)$, for some $n\in\NN$; and
$\textit{max}_j(x)$ states that 
\begin{enumerate}[\ 1.]
 \item
   the (reverse) binary representation of $n$, 
   i.e., the $\set{0,1}$-string built from the letters in $u$ 
   that occur directly to the right of letters $\TagE{j{-}1}$, is of the form
   $011\cdots 11$, and 
 \item
   the highest bit of $n$ has the number $\Tower(j{-}1){-}1$, i.e.,
   the rightmost position $y$ in $u$ that carries the letter $\Tag{j{-}1}$ satisfies the
   formula $\max_{j-1}(y)$. 
\end{enumerate}
Obviously, $\textit{max}_j(x)$ can be formalised in $\FO(\tau_h)$ by a formula of 
size $\bigO(j)$.
Finally, this completes Step~1.
\medskip\\
%
\underline{\emph{Step 2:}} \ 
A string $w$ over $\Sigma^{\bullet}_h$ belongs to the language $(v_h)^+$ if, and only if,
all the following conditions are satisfied:
\begin{enumerate}[\ 1.]
 \item
   $w$ satisfies \,$\textit{ok}_1\und\cdots\und\textit{ok}_h$, 
 \item
   the first position in $w$ carries the letter $\Tag{h{+}1}$, the last position in
   $w$ carries the letter $\TagE{h{+}1}$, 
   the letter $\Tag{h{+}1}$ occurs at a position $x>1$ iff position $x{-}1$ carries
   the letter $\TagE{h{+}1}$,
   and the letter $\bullet$ occurs at a
   position $x$ iff position $x{-}1$ carries the letter $\TagE{h}$ and
   position $x{+}1$ carries the letter $\Tag{h}$ or $\TagE{h{+}1}$,
 \item
   whenever $x$ (respectively, $y$) is a position in $w$ that carries the letter 
   $\Tag{h{+}1}$ (respectively, $\TagE{h{+}1}$) the following is true:
   There is a position $y$ to the right of $x$ that carries
   the letter $\TagE{h{+}1}$ (respectively, a position $x$ to the left of $y$ that 
   carries the letter $\Tag{h{+}1}$), such that the substring $u$ of $w$ that 
   starts at position $x$ and ends at position $y$ is of the form $v_h$, i.e.,
   \begin{enumerate}[$\star$]
    \item
      the letters $\Tag{h{+}1}$ and $\TagE{h{+}1}$ only occur at the first and
      the last position of~$u$, 
    \item
      the prefix of length 3 of $u$ is of the form 
      \,$\Tag{h{+}1}\Tag{h}\TagE{h}$, and the suffix of length 3 of $u$ is of the
      form \,$\TagE{h}\bullet\TagE{h{+}1}$, 
    \item
      whenever $x'$ and $y'$ are positions in $u$ carrying the letter $\Tag{h}$ such
      that $x'<y'$ and no position between $x'$ and $y'$ carries the letter $\Tag{h}$,
      the formula $\textit{inc}_h(x',y')$ from 
      Lemma~\ref{lemma:inc-h} is satisfied,
    \item
      the rightmost position $x''$ in $u$ that carries the letter $\Tag{h}$ satisfies
      the formula $\textit{max}_h(x'')$.
   \end{enumerate}
\end{enumerate}
Using the formulas constructed in Step~1 and the preceding lemmas, it is 
straightforward to see that this can be formalised by an $\FO(\tau_h)$-formula
$\varphi_{(v_h)^+}$ of size $\bigO(h^2)$. 
This finally completes the proof of Lemma~\ref{lemma:FO-def-of-vhPlus}.
\Qed
\end{proof}
\\
\parno
Finally, we are ready for the
\\
\parno
\begin{proofc}{of Lemma~\ref{lemma:MSO-definability}}\mbox{}\\
To determine whether an input string $w$ is indeed the string 
$w_h$, one can proceed as follows: First, we make sure that
the underlying string $w$ belongs to $(v_h)^+$ via the $\FO(\tau_h)$-formula
$\varphi_{(v_h)^+}$ of Lemma~\ref{lemma:FO-def-of-vhPlus}. 
Afterwards we, in particular, know that in each $\Tag{h{+}1}\cdots\TagE{h{+}1}$-block
is of the form $v_h$ and therefore contains exactly $H$ positions that carry the 
letter $\bullet$.
Now, to each $\bullet$-position in $w$ we assign a letter from $\set{0,1}$ in such
a way that the $\set{0,1}$-string built from these assignments is an 
\emph{$H$-numbering}, i.e., of one of the following forms:
\begin{enumerate}[\ 1.]
\item
   $\BIN_H(0)\,\BIN_H(1)\,\BIN_H(2)\cdots\BIN_H(n)$, \quad for some $n<2^H$, 
\item
   $\BIN_H(0)\,\BIN_H(1)\cdots\BIN_H(2^H{-}1)\,\Big(\BIN_H(0)^m\Big)$, \quad for some $m\geq 0$.
\end{enumerate} 
Here, $\BIN_H(n)$ denotes the reverse binary representation of length $H$ of the 
number $n<2^H$. For example, $\BIN_4(2) = 0100$ and $\BIN_4(5) = 1010$.
Of course, $w$ is the string $w_h$, i.e., consists of exactly $2^H$ copies of $v_h$,
if and only if the $H$-numbering's assignments in the rightmost copy of $v_h$ 
form the string $\BIN_H(2^H-1)$, i.e., if and only if every 
$\bullet$-position in this copy of $v_h$ was assigned the letter 1. \\
One way of assigning letters from $\set{0,1}$ to the $\bullet$-positions in $w$ is
by choosing a set $X$ of $\bullet$-positions with the intended meaning that a
$\bullet$-position $x$ is assigned the letter 1 if $x\in X$ and the letter 0 if
$x\not\in X$.
\par
Using the $\FO(\tau_h)$-formulas $\textit{equal}_h$ 
of Lemma~\ref{lemma:equal-h}
and $\varphi_{(v_h)^+}$ of Lemma~\ref{lemma:FO-def-of-vhPlus}, 
it is straightforward to construct the desired 
$\exists X\,\FO(\tau_h)$-formula $\Phi_h$ of size $\bigO(h)$.
\\
This completes the proof of Lemma~\ref{lemma:MSO-definability}.
\Qed%
\end{proofc}
\\
\parno
As a consequence of 
Lemma~\ref{lemma:MSO-definability} 
and Lemma~\ref{lemma:FOkleiner} one obtains a non-elementary succinctness gap between
$\FO$ and $\MSO$ on the class of linear orders:
\begin{theorem}\label{theorem:FOvsMSO-UnaryAlphabet}\mbox{}\\
The $\FO(<)$-expressible fragment of $\MonSigma_1^1$ is 
not $\Tower\big(o(\sqrt{m})\big)$-succinct
in $\FO(<)$ on the class of linear orders.\Fertig
\end{theorem}%
\begin{proof} 
Recall that, for every $N\in\NN$, $\A_N$ denotes the linear order with universe
$\set{0,\twodots,N}$.
\par
For every $h\in\NN$ let \,$\ell(h) \deff |w_h| -1$, \,where $|w_h|$ denotes the length of the string $w_h$.
We say that a sentence $\chi$ \emph{defines} the linear order $\A_{\ell(h)}$ if, and only if,
$\A_{\ell(h)}$ is the unique structure in $\setc{\A_N}{N\in\NN}$ that satisfies $\chi$.
For every $h\in\NN$ we show the following:
\begin{enumerate}[\emph{(a)}]
\item
  Every $\FO(<,\Succ,\Min,\Max)$-sentence $\psi_h$ that defines $\A_{\ell(h)}$
  has size $\size{\psi_h}\geq \Tower(h)$.
\item
  There is a $\MonSigma_1^1(<)$-sentence $\Psi_h$ of size $\size{\Psi_h}=\bigO(h^2)$  that
  defines $\A_{\ell(h)}$. 
\end{enumerate}
\textit{Ad (a):}\\
Since $w_h$ consists of $2^{\Tower(h)}$ copies of $v_h$, we know that 
$\ell(h)\geq 2^{\Tower(h)}$. Therefore, every $\FO(<)$-sentence $\psi_h$ that defines $\A_{\ell(h)}$ has
quantifier depth, and therefore size, at least $\Tower(h)$ (cf., Lemma~\ref{lemma:FOkleiner}).
\medskip\\
\textit{Ad (b):}\\
Let $\exists X\,\varphi_h$ be the $\exists X\,\FO(\tau_h)$-sentence obtained from Lemma~\ref{lemma:MSO-definability}.
It is straightforward to formulate an $\FO(\tau_h)$-sentence $\xi_h$ of size $\bigO(h^2)$ which 
expresses that every element in the underlying structure's universe belongs to exactly one of the sets
$P_{\sigma}$, for $\sigma\in\Sigma_h^{\bullet}$.
\\ 
The $\MonSigma_1^1(<)$-sentence
\begin{eqnarray*}
 \Psi_h & \ \deff \ & 
 \big(\exists P_{\sigma}\big)_{\sigma\in\Sigma_h^{\bullet}}\ \ \exists\, X \ \ \
   (\,\xi_h \und \varphi_h\,)
\end{eqnarray*}
expresses that the nodes of the underlying linear order can be labelled with letters in
$\Sigma_h^{\bullet}$ in such a way that one obtains the string $w_h$. Such a labeling 
is possible if, and only if,
the linear order has length $|w_h|$. I.e., $\Psi_h$ \emph{defines} $\A_{\ell(h)}$. 
Furthermore, $\size{\Psi_h}=\bigO(h^2)$, because $\size{\xi_h}=\bigO(h^2)$ and 
$\size{\varphi_h}=\bigO(h^2)$.
\\
This completes the proof of Theorem~\ref{theorem:FOvsMSO-UnaryAlphabet}
\Qed
\end{proof}%
\\
\parno
Let us remark that by modifying the proof of the above result, one 
can also show that the $\FO(<)$-expressible fragment of 
\emph{monadic least fixed point logic}, $\MLFP$, is 
non-elementarily more succinct than
$\FO(<)$ on the class of linear orders.
\medskip

\section{Conclusion}\label{section:Conclusion}%
%
%
%
Our main technical result is a lower bound on the size of a 3-variable formula
defining a linear order of a given size. 
We introduced a new technique
based on Adler-Immerman games that might be also useful in other situations.
A lot of questions remain open, let us just mention a few here:
\begin{enumerate}[$\bullet$]
\item
Is first-order logic on linear orders $\text{poly}(m)$-succinct in its
4-variable fragment, or is there an exponential gap?
\item
As a next step, it would be interesting to study the succinctness of the
finite-variable fragments on strings, that is, linear orders with additional
unary relation symbols. It is known that on finite strings, the 3-variable
fragment of first-order logic has the same expressive power as full
first-order logic. Our results show that there is an at least exponential
succinctness gap between the 3-variable and the 4-variable fragment. We do not
know, however, if this gap is only singly exponential or larger, and we also
do not know what happens beyond 4 variables.
\item
Another interesting task is to study the succinctness of various extensions of
(finite variable fragments of) first-order logic by transitive closure operators.
\item
It also remains to be investigated if our results can possibly help to settle
the long standing open problem of whether the 3-variable and 4-variable
fragments of first-order logic have the same expressive power on the
class of all ordered finite structures. 
\end{enumerate}%
Finally, let us express our hope that techniques for proving lower bounds on
succinctness will further improve in the future so that simple results such as
ours will have simple proofs!
\medskip
%
%
%

\end{document}